\documentclass[twocolumn,preprintnumbers,showpacs,superscriptaddress,%
nofootinbib,amsmath,amssymb,prd,aps,floatfix]{revtex4-1}

\usepackage{graphicx}
\usepackage[usenames,dvipsnames,svgnames,table]{xcolor}

\begin{document}

\preprint{arXiv:1512.09375 [gr-qc]}

\title{Cosmological stealths with nonconformal couplings}

\author{Eloy Ay\'on-Beato}
\email{ayon-beato-at-fis.cinvestav.mx}
\affiliation{Departamento~de~F\'{\i}sica,~CINVESTAV-IPN,%
~Apartado.~Postal~14-740,~07000,~M\'exico~D.F.,~M\'exico}

\author{P.\ Isaac Ram\'irez-Baca}
\email{pramirez-at-fis.cinvestav.mx}
\affiliation{Departamento~de~F\'{\i}sica,~CINVESTAV-IPN,%
~Apartado.~Postal~14-740,~07000,~M\'exico~D.F.,~Mexico}

\author{C\'esar A.\ Terrero-Escalante}
\email{cterrero-at-ucol.mx}
\affiliation{Facultad de Ciencias,~Universidad de Colima,%
~C.P.~28045~Colima,~Colima,~Mexico.}

\begin{abstract}
In this paper we reconsider the existence of stealth fields during the
evolution of our Universe by admitting more realistic nonminimal couplings
to gravity than the conformal one. In the framework of the FRW cosmology we
found that inhomogeneous stealths, as those enabled in the conformal case,
are only allowed in de Sitter backgrounds. In particular, it is shown that
the homogeneous stealths resulting from nonconformal couplings can coexist
with each kind of matter dominant in the several phases characterizing the
evolution of our Universe according to the $\Lambda$CDM model.	
\end{abstract}

\maketitle


\section{Introduction\label{sec:intro}}

The standard cosmological model describes the evolution of an isotropic and
homogeneous universe as required by the cosmological principle, that is of a
spacetime described by the Friedmann-Robertson-Walker (FRW) metric
\begin{equation}\label{eq:FRW}
ds^2=a(\tau)^2\left(-d\tau^2+\frac{dr^2}{1-kr^2}
+r^2\left(d\theta^2+\sin^2\!\theta\,{d}\phi^2\right)\right),
\end{equation}
where $\tau$ is the conformal time and $a(\tau)$ is the scale factor. The
constant $k$ determines the spatial topology of the Universe, with
$k=-1,0,+1$ corresponding to open, flat and closed universes. For a given
matter content $\varphi_{\text{m}}$, metric \eqref{eq:FRW} extremizes the
action
\begin{equation}\label{eq:action}
S[g,\varphi_{\text{m}}]=\int{d}^4x\sqrt{-g}\left(
\frac1{2\kappa}(R-2\Lambda)+L_{\text{m}}\right),
\end{equation}
by solving Einstein equations
\begin{equation}\label{eq:EEqs}
G_{\mu\nu}+\Lambda g_{\mu\nu}-\kappa{T}_{\mu\nu}=0\, ,
\end{equation}
where $\kappa$ is the Einstein constant and $\Lambda$ is the cosmological
constant. A consistent picture of the cosmological history, supported by the
observational data \cite{Bernabei:2014ema,
Suzuki:2011hu,Dawson:2012va,Aghanim:2015xee}, emerged from the corresponding
solutions of \eqref{eq:EEqs}. Known as the Big Bang, it describes an early
very flat universe filled with a quark-gluon plasma which started to expand
and to get colder, giving rise to many phase transitions in the matter
content. This picture involves a mixture of fluids, however, since the
density of each fluid varies differently with the expansion, there are long
periods of time where the energy of a given fluid is dominant while the other
ones are subdominant. This way, the early radiation dominated phase was
replaced, about $350000$ years later, by an intermediate era dominated by
pressureless matter (baryonic and of an unknown kind coined dark). By the end
of the last century it was empirically established that very recently our
observable universe became dominated by some energy (also of an unknown kind
and also coined dark) doing a positive work which induces accelerated
expansion. The simplest candidate for this dark energy is a positive
cosmological constant.

All of the above mentioned cosmological phases can be modeled with barotropic
perfect fluids given by an equation of state $p=w\rho$, where $p$ and $\rho$
stand, correspondingly, for the pressure and energy density of the fluid,
while $w$ is the constant of state. In absence of a cosmological constant,
the solution of Einstein equations \eqref{eq:EEqs} for the flat FRW universe
\eqref{eq:FRW} filled with a barotropic fluid corresponds to a power law (see
Appendix \ref{app:SF_BPF})
\begin{equation}
a(\tau)=a_b\tau^{b},\qquad b\equiv\frac{2}{1+3w}.
\label{eq:PowLawSF}
\end{equation}
For radiation, $w=1/3$, the power is $b=1$, while for baryonic and dark
pressureless matter, $w=0$, it is $b=2$. Even more, the case of a universe
dominated by the cosmological constant can also be described by
\eqref{eq:PowLawSF} with a constant of state $w=-1$. As it was mentioned, the
corresponding solution with $b=-1$ is accelerating and is known as the de
Sitter spacetime.

Since the described background is homogeneous and isotropic, the picture
briefly drawn cannot explain the existence of cosmic large scales structures, 	
which are the consequence of the growth of primordial density fluctuations.
Amongst some other problems, it neither explains why our Universe looks so
flat, given that, with a positive constant of state, the flat case is an
unstable equilibrium of the cosmological dynamics
\cite{Liddle:2000cg,Mukhanov:2005sc}. Up to date, the best available solution
to these problems is called cosmological inflation
\cite{Liddle:2000cg,Mukhanov:2005sc}. It also describes an accelerated
expansion, but at very early times, even before the formation of the
quark-gluon plasma. The cosmological constant is not a good candidate for the
inflationary matter because the de Sitter universe is unstable in the sense
that it never ends to accelerate, thus not giving place to the rise and
expansion of the quark-gluon plasma. Therefore, perhaps the simplest scenario
is that of the energy density in the very early universe being dominated by
the potential energy of a single real scalar field $\phi$ called inflaton.
Then, the quantum fluctuations of the inflaton served as the seeds for the
formation of large scale structure and of the anisotropies in the cosmic
microwave background (CMB). An interesting model is power-law inflation for
potential $V(\phi)\propto\exp(-\sqrt{2\kappa (b+1)/b}\,\phi)$ and scale
factor \eqref{eq:PowLawSF} with $b<-1$ \cite{Lucchin:1984yf}. It has several
attractive features concerning theory and observations
\cite{Liddle:2000cg,Mukhanov:2005sc}. First, the tree expansions often used
for scalar field potentials arising in quantum field theory are just
deviations from the truncated Taylor series of the exponential function.
Second, the exponential potential is the only inflationary model yielding
exact power-law spectra \cite{Abbott:1984fp}, and because the departure of
the spectra amplitudes from scale invariance is predicted to be very small,
the power-law parametrization of the primordial spectra is often used while
analyzing the CMB data \cite{Aghanim:2015xee}.

This whole description of the evolution of our Universe is known as the $\Lambda$ cold dark matter
($\Lambda$CDM) model. Data from highly accurate satellite and ground-based
observations strongly support this model as the best theoretical framework
\cite{Aghanim:2015xee}. Nevertheless, in spite of the success of the $\Lambda$CDM model, there is still a large number of open questions in cosmology,
perhaps the more relevant being what is the nature of most of the matter
content of our Universe. We do not know what are the dark energy, the dark
matter or the precise inflationary matter. Typically, candidates for these
kinds of matter are accepted or discarded considering the impact they have on
the behavior of the scale factor, which in turn determines the values of most
cosmological observables. Additionally, a less known kind of exotic matter
can also be included in the $\Lambda$CDM model. It has been shown that
General Relativity allows for special nontrivial matter configurations with
no backreaction on the gravitational field, i.e.\ whose existence does not
affect the evolution of the underlying spacetime. Scalar fields with this
property were first found for the static BTZ black hole
\cite{AyonBeato:2004ig} and later for the Minkowski flat space
\cite{AyonBeato:2005tu}, (A)dS space \cite{AyonBeato:SAdS} and other
holographic backgrounds \cite{Ayon-Beato:2015qfa}. They were coined
gravitational \emph{stealths}. The existence of stealths for the de~Sitter
(dS) cosmology was studied in Ref.~\cite{Banerjee:2006pr}. The relevance they
have for the probability of nucleation of such universes was emphasized in
Ref.~\cite{Maeda:2012tu}.

Moreover, in Ref.~\cite{Ayon-Beato:2013bsa} it was shown that not only
de~Sitter universes, but \emph{any} homogeneous and isotropic universe,
independently of its spatial topology and matter content, allows for the
presence of a conformal stealth which evolves along with the Universe without
exhibiting its gravitational fingerprints. This result can be understood as a
mapping of the previously known stealth configurations of Minkowski spacetime
into stealths in FRW spacetimes. The corresponding map was found using the
conformal symmetry of the stealth action, as well as the conformal flatness
of FRW spacetime which is a consequence of its homogeneity and isotropy.
Nevertheless, as it was already mentioned, the later are just approximated
symmetries of the actual Universe where stars, galaxies, clusters of
galaxies, etc.\ are present. Therefore, many important observables of the
successful $\Lambda$CDM model are related to density perturbations which grew
into this cosmic large scale structure. The backreaction of these density
perturbations, on the one hand breaks the above mentioned symmetries of the
FRW spacetimes, spoiling its conformal flatness, and on the other hand they
also induce perturbations on the coexisting conformal stealth. These last
perturbations differ from the stealth in nature because they are necessarily
coupled both to the matter sources, which were never conformally invariant,
and to the perturbed spacetime which is not longer conformally flat.
Consequently, conformal symmetry is irrelevant at the perturbative level
defining the $\Lambda$CDM model. Hence, constraining the stealths of
cosmology to being conformally coupled is an unnecessary restriction which
can severely limit the analysis of more realistic scenarios. For this reason,
the main motivation of this paper is to explore the existence of stealth
configurations in FRW spacetimes for general nonminimal couplings, where no
obvious conformal arguments can be used.

In the next section we state the problem for a general nonminimal coupling
parameter $\xi$ and derive the relevant expressions. One main conclusion is
that going beyond the conformal coupling, $\xi\neq1/6$, necessarily implies
that a generic cosmology only allows homogeneous stealths. In contrast with
the conformal case, $\xi=1/6$, the only exception are just the de Sitter
cosmologies. In Sec.~\ref{sec:LCDM} we first show the existence of a class of
stealths during a big bang era for generic values of $\xi$. Later we present
some examples of stealth potentials for different couplings during those eras
of the $\Lambda$CDM model when the matter content was dominated by the energy
of a single fluid as those yielding \eqref{eq:PowLawSF}. Next, in
Sec.~\ref{sec:dS} the inhomogeneous stealths of the de Sitter cosmologies are
analyzed in detail for generic values of the nonminimal coupling parameter.
The special value of the coupling $\xi=1/4$ is studied in
Sec.~\ref{sec:spcase14}. For homogeneous stealths the nonminimal coupling to
gravity can be reinterpreted as contributing to the self-interaction of the
field; the arising of this effective self-interaction potentials is
discussed in Sec.~\ref{sec:Ueff}. In the last section we briefly summarize
our conclusions. We also include some appendixes with relevant details.

\section{Stealths with nonconformal couplings\label{sec:NCC}}

Let us start here by stating the problem of finding stealth scalar fields not
necessarily conformally coupled to the FRW spacetime \eqref{eq:FRW}. With
this aim we supplement action \eqref{eq:action} with the one describing a
self-interacting scalar field $\Psi$ nonminimally coupled to gravity
\begin{equation}
S[g,\varphi_{\mathrm{m}}]-\int{d}^4x\sqrt{-g}\left(
\frac12\partial_{\mu}\Psi\partial^{\mu}\Psi+\frac{\xi}{2} R\Psi^2+U(\Psi)
\right),
\label{eq:action+s}
\end{equation}
where $\xi$ is the nonminimal coupling parameter, and the special value
$\xi=1/6$ is the one describing the conformal case. Next, we demand the first
term to be again extremized by solving Einstein equations \eqref{eq:EEqs} for
the same matter content $\varphi_{\mathrm{m}}$. This necessarily implies that
the second term does not contribute to the extrema, i.e.\ the corresponding
energy-momentum tensor must vanish on the background,
\begin{align}
0=T^{\mathrm{s}}_{\mu\nu}={}&\partial_{\mu}\Psi\partial_{\nu}\Psi
-g_{\mu\nu}\left(\frac{1}{2}
\partial_{\alpha}\Psi\partial^{\alpha}\Psi+U(\Psi)\right)
\nonumber\\
&+\xi\left(g_{\mu\nu}\square-\nabla_{\mu}\nabla_{\nu}
+G_{\mu\nu}\right)\Psi^2 \,,\label{eq:En-Mom-ten}
\end{align}
and any nontrivial solution $\Psi$ to the above constraint defines a stealth
configuration.

In order to find these nontrivial solutions on a given FRW background with
scale factor $a$ it is useful to redefine the field as
\begin{equation}\label{eq:Psi2sigma}
\Psi=\frac{1}{\sqrt{\kappa}\left(a\sigma\right)^\frac{2\xi}{1-4\xi}},
\end{equation}
where the function $\sigma=\sigma(x^\mu)$ inherits the full spacetime
dependence of the scalar field. This redefinition is chosen because it allows
to follow the general method used in \cite{Ayon-Beato:2013bsa} for the
conformal case. From now on the nonminimal coupling parameter $\xi$ is taken
to be nonvanishing (because there is no stealth for minimal coupling $\xi=0$)
and different from the value $\xi=1/4$, which will be studied in
Sec.~\ref{sec:spcase14}.

Hence, following \cite{Ayon-Beato:2013bsa} we conclude, by solving the
off-diagonal components of the stealth constraint \eqref{eq:En-Mom-ten},
that the redefined field is again separable according to
\begin{equation}\label{eq:sigma_s}
\sigma(x^\mu)=T(\tau)+R(r)+r\left[\Theta(\theta)+\sin(\theta)\Phi(\phi)\right].
\end{equation}
This separation is defined up to a residual symmetry; homogeneous terms in
$\Phi$, $\Theta$, and $R$ can be compensated by a sinusoidal dependence in
$\Theta$, a linear one in $R$, and another homogeneous term in $T$,
respectively.

Now, by taking differences of the diagonal components of
\eqref{eq:En-Mom-ten}, it is found that as in \cite{Ayon-Beato:2013bsa}
\begin{subequations}\label{eq:RTP}
\begin{align}
\Phi(\phi) &= A_1\cos\phi+A_2\sin\phi,\\
\Theta(\theta) &= A_3\cos\theta, \\
R(r) &= \left\{
\begin{array}{ll}
B_{-}\sqrt{1-kr^2}, & k\ne0,\\ \\
\frac12\alpha{r}^2, & k=0,
\end{array}\right.
\end{align}
\end{subequations}
where the above mentioned residual symmetry has been taken into account.
Here, $A_i$ ($i=1,2,3$), $\alpha$ and $B_{-}$ are integration constants whose
dimensions are inverse length.

The remaining difference of diagonal components is
\begin{align}
0&=\frac{(1-4\xi)a^2\sigma}{4\xi^2\Psi^2}\left(T_{~\theta}^{\mathrm{s}~\theta}
-T_{~\tau}^{\mathrm{s}~\tau}\right)\nonumber\\
&=
\begin{cases}
T''+kT+\dfrac3\xi\left(\xi-\dfrac16\right)
\left(\mathcal{H}'-\mathcal{H}^2-k\right)\sigma, & k\ne0,\\ \\
T'' +\alpha+\dfrac3\xi
\left(\xi-\dfrac16\right)\left(\mathcal{H}'-\mathcal{H}^2\right)\sigma, & k=0,
\end{cases}
\label{eq:T_te^te-T_t^t}
\end{align}
where $\mathcal{H}\equiv a'/a$ is the conformal Hubble parameter and the
prime denotes derivatives with respect to conformal time. Is here where the
first differences with respect to the approach of
Ref.~\cite{Ayon-Beato:2013bsa} appear. In that reference, the inhomogeneous
$\sigma$ contributions of conditions \eqref{eq:T_te^te-T_t^t} do not appear
because the studied coupling had the conformal value $\xi=1/6$. This allowed
us to find exactly the time dependent contribution $T$, regardless of the
scale factor $a$ and the corresponding spatial dependence of the stealth.
This is no longer valid for the general nonminimal couplings, $\xi\neq1/6$,
considered here. As can be appreciated from \eqref{eq:T_te^te-T_t^t}, the
stealth must be necessarily homogeneous, unless the parentheses including the
conformal Hubble parameter vanishes, i.e.
\begin{equation}\label{eq:dSConst}
\mathcal{H}'-\mathcal{H}^2-k=0.
\end{equation}
Only when the above constraint is fulfilled inhomogeneous configurations
having $\sigma(x^\mu)\neq T(\tau)$ are allowed, independently of the value
taken by the nonminimal coupling parameter. Constraint \eqref{eq:dSConst} has
a single solution which is precisely the de Sitter universe
(Appendix \ref{app:appdS}), whose inhomogeneous stealths will be studied in
Sec.~\ref{sec:dS}. In other words, it follows from \eqref{eq:T_te^te-T_t^t}
that the only cosmologies allowing inhomogeneous stealths for a generic value
of the nonminimal coupling parameter $\xi\neq1/6$ are the de Sitter ones. For
any other cosmology it is necessarily satisfied that $\sigma(x^\mu)=T(\tau)$,
i.e.\ the resulting stealths become homogeneous, which according to
\eqref{eq:sigma_s} only occurs when all the integration constants appearing
in solution \eqref{eq:RTP} vanish. Summarizing, in order to completely
determine the stealth from expression \eqref{eq:Psi2sigma} for a nonconformal
coupling and a given scale factor (defining a cosmology different from the de
Sitter one), it remains to find the extra time dependence from
Eqs.~\eqref{eq:T_te^te-T_t^t}. For arbitrary backgrounds, no general
closed-form solution exists for these equations, i.e.\ it will depend on the
specific form of the scale factor.

After solving \eqref{eq:T_te^te-T_t^t}, we would have satisfied all the
off-diagonal conditions of the stealth constraint \eqref{eq:En-Mom-ten}, as
well as all the differences between the diagonal ones. Hence, only one last
condition needs to be satisfied, which fixes the unique potential defining
the self-interacting behavior of the stealth. For a generic cosmology this
condition is
\begin{align}\label{eq:T00_const}
0=-T_{~\tau}^{\mathrm{s}~\tau}=
{}&U(\Psi)-\frac{\Psi^2}{2a^2}\biggl\{6\xi\left[(6\xi-1)\mathcal{H}^2-k\right]
\nonumber\\
&\qquad\qquad\qquad-\left[\left(\ln{a^{6\xi}\Psi}\right)'\right]^2\biggr\}.
\end{align}
In the next section we start by finding the explicit form of the nonconformal
stealths for the scale factors which describe the different eras of
$\Lambda$CDM cosmology.

\section{Stealths in $\Lambda$CDM Cosmology\label{sec:LCDM}}

Perhaps the more interesting solution to the problem stated in the previous
section is a stealth witnessing the whole cosmological evolution as described
by the $\Lambda$CDM model. In principle, the solution on the related
background can be found by numerically integrating \eqref{eq:T_te^te-T_t^t},
however, for a better understanding of the main features of this universal
stealth it is useful to do the analysis for each of the eras comprising the
$\Lambda$CDM model.

With this aim we start by analyzing the Big Bang era. As it was already
mentioned, it describes the evolution of the mixture of two kinds of
barotropic perfect fluids: radiation and pressureless matter. Let
$\rho_{\mathrm{\nu}}$ and $\rho_{\mathrm{m}}$ be the densities for radiation
and matter respectively, then (as is explicitly shown in
Appendix \ref{app:SF_mix_BPF}) one can solve for the scale factor as
\begin{subequations}
\begin{equation}
a(\tau)=a_1\tau+a_2 \tau^2,
\label{eq:ScaFacRadMatt}
\end{equation}
where
\[
a_1=\sqrt{\frac{3}{\kappa\rho_{\mathrm{\nu}}}}, \quad
a_2=2\sqrt{\frac{3}{\kappa\rho_{\mathrm{m}}	}},
\]
\end{subequations}
and the powers are fixed according to \eqref{eq:PowLawSF}. With this scale
factor, the homogeneous solution of Eq.~\eqref{eq:T_te^te-T_t^t} for $k=0$ is
given in terms of the hypergeometric function
\begin{subequations}\label{eq:cumber}
\begin{align}
T(\tau)={}&a(\tau)^{\frac12-\beta_1}\nonumber\\
&\times\!\biggl[C_1\;_{2}F_1\!\!\left(\gamma_{-}-2\beta_1,\gamma_{+}-2\beta_1
;1-2\beta_1;-\frac{a_2}{a_1}\tau\right)\nonumber\\
&+C_2\tau^{2\beta_1}\,_{2}F_1\!\!\left(\gamma_{-},\gamma_{+};1+2\beta_1;
-\frac{a_2\tau}{a_1}\right)\biggr],
\end{align}
where
\begin{equation}
\beta_1=\sqrt{\frac{25}{4}-\frac{1}{\xi}},\quad
\gamma_{\pm}=\frac12\pm\sqrt{\frac{73}{4}-\frac{3}{\xi}},
\end{equation}
\end{subequations}
and $C_1$, $C_2$ are integration constants. Recall that to find the
self-interaction potential we need to solve the constraint
\eqref{eq:T00_const}. The usual way to deal with this problem is to write the
explicit dependence of both the stealth and the scale factor on the
coordinates (in this case $\tau$) and then use the inverse relation between
the stealth and the coordinates to exhibit the explicit dependence $U(\Psi)$.
However, relation \eqref{eq:Psi2sigma} is not invertible in general for an
arbitrary nonminimal coupling $\xi$. This is the case for the scale factor
\eqref{eq:ScaFacRadMatt}.

In spite of the above difficulty we can outline the main properties of these
stealths if we use the fact that initially ($\tau<<1$) the first contribution
in the scale factor \eqref{eq:ScaFacRadMatt} dominates, indicating a
radiation phase, and lately ($\tau>>1$) the leading term is the second one,
corresponding to a matter dominated phase. This way in each case the scale
factor is reduced to just a power-law. Even more, as mentioned in the
Introduction, another important phase of the $\Lambda$CDM cosmology which is
also described by a power-law is inflation. Taking into account all these
reasons, we will consider a power-law evolution in flat universes as
described by the scale factor \eqref{eq:PowLawSF}. After introducing this
expression in the stealth condition \eqref{eq:T_te^te-T_t^t} we obtain the
following differential equation
\begin{equation}
\tau^2T''-\frac{3b(b+1)(\xi-1/6)}{\xi}T=0.
\label{eq:TEqBaroF}
\end{equation}
This is an Euler equation and their linearly independent solutions are also
power-laws, $T\propto\tau^\gamma$, whose exponents $\gamma$ are determined
from the corresponding characteristic polynomial
\begin{equation}
\gamma^2-\gamma-\frac{3b(b+1)(\xi-1/6)}{\xi}=0,
\end{equation}
solved by
\begin{subequations}
\begin{align}
\gamma_{\pm}&=\frac12\pm\sqrt{\Delta}\,,\\
\Delta&=\frac{b(b+1)}{2}\left(\frac{1}{\xi_b}-\frac{1}{\xi}\right),
\label{eq:discriminant}\\
\xi_b&=\frac{1}{6+\frac{1}{2b(b+1)}}.
	\end{align}
\end{subequations}%
\begin{figure}
\includegraphics[width=\columnwidth]{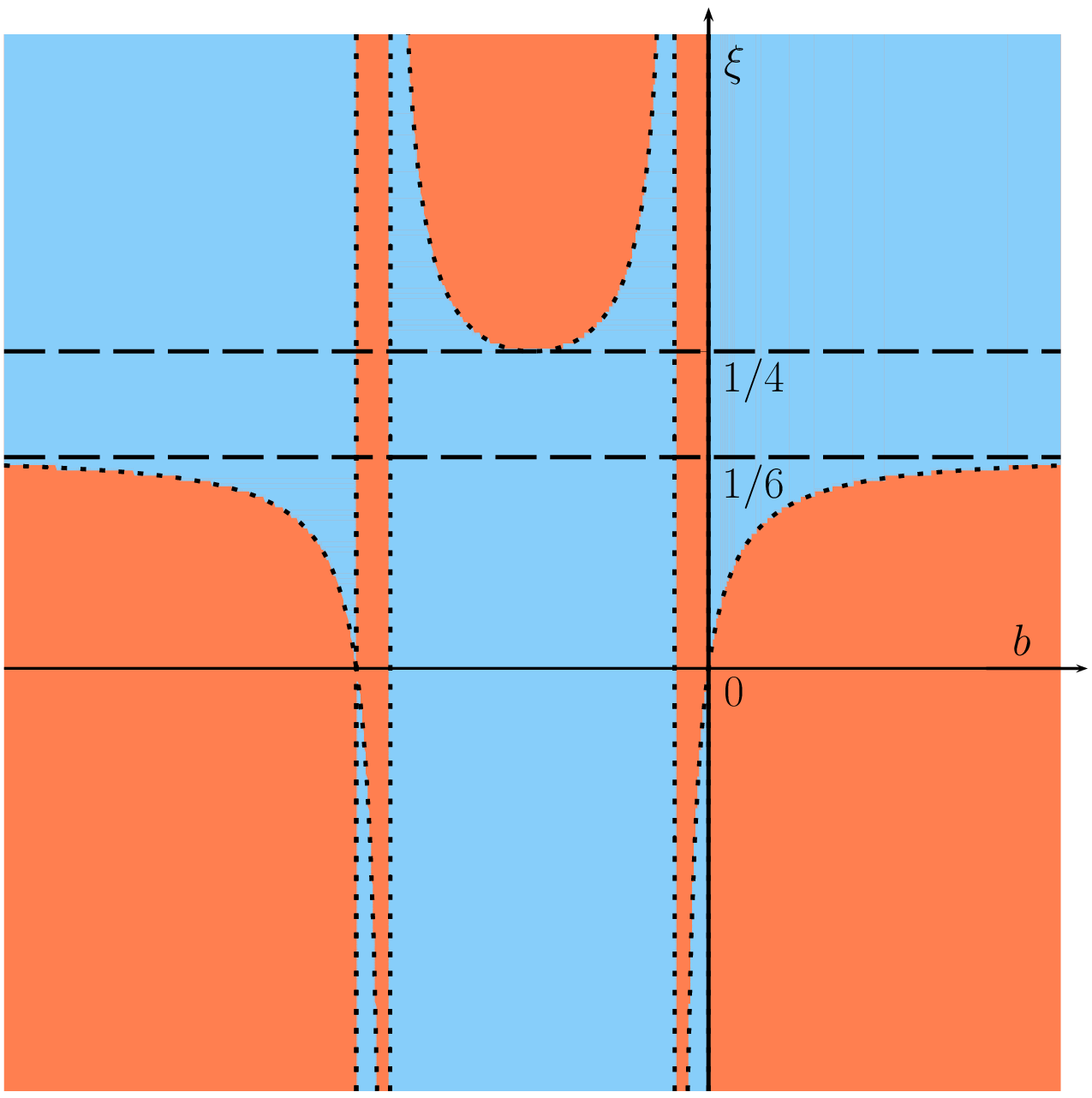}
\caption{Regions where the discriminant $\Delta$ is positive (light blue) and
negative (salmon).}
\label{fig:ExclusionZone}
\end{figure}%
Therefore, the precise functional form of the solutions of the Euler equation
\eqref{eq:TEqBaroF} depends on the sign of the discriminant $\Delta$. These
signs can be appreciated in Fig.~\ref{fig:ExclusionZone}, where the blue
regions denote $\Delta>0$, the salmon ones $\Delta<0$ and consequently
$\Delta=0$ are just the boundaries between these regions. We analyze each of
these cases in the following subsections.

\subsection{Solutions with $\Delta>0$\label{ssec:SPD}}

The discriminant \eqref{eq:discriminant} is positive for two general cases:
the first, when $-1<b<0$ with $\xi<\xi_b$, and the second, for $b>0$ or
$b<-1$ with $\xi>\xi_b$ (see Fig.~\ref{fig:ExclusionZone}). In those cases
the general solution of the Euler equation \eqref{eq:TEqBaroF} is given by
\begin{subequations}
\begin{equation}\label{eq:EulerEqSln}
T(\tau)=A_0 \tau^{\frac12+\beta}+\sigma_0 \tau^{\frac12-\beta},
\end{equation}
where $A_0$ and $\sigma_0$ are integration constants and
\begin{equation}
\beta=\sqrt{\Delta}\,.\label{eq:beta}\\
\end{equation}
\end{subequations}
Using this solution the stealth field \eqref{eq:Psi2sigma} can be expressed
as
\begin{subequations}\label{eq:hsNonOscSol}
\begin{equation}\label{eq:hsNonOscSolA}
\Psi(\tau)=\Psi_0\left[\left(\frac{\tau}{\tau_0}\right)^{b + \frac12+\beta}
+\left(\frac{\tau}{\tau_0}\right)^{b+\frac12-\beta}
\right]^{\frac{-2\xi}{1-4\xi}},
\end{equation}
where the new constants are related to the old ones by
\begin{align}
\tau_0=&\left(\frac{\sigma_0}{A_0}\right)^\frac{1}{2\beta},\\
\Psi_0=&\frac{1}{\sqrt\kappa}\left[a_0
\sigma_0^{\frac{b+\frac12+\beta}{2\beta}}
A_0^{-\frac{b+\frac12-\beta}{2\beta}}\right]^{\frac{-2\xi}{1-4\xi}}.
\end{align}
\end{subequations}
In general, expression \eqref{eq:hsNonOscSolA} is not invertible and the
corresponding self-interaction potential must be parametrically given by
\[
U(\Psi) =
\left\{
	\begin{array}{ll}
		\Psi(\tau), \\
		U(\tau),
	\end{array}
\right.
\label{eq:ParamPot}
\]
where $U(\tau)$ is given by \eqref{eq:T00_const}.

It is worthy to notice that the relation between $\tau$ and $\Psi$ can be
inverted for some specific values of the powers of $\tau$ that appear in
\eqref{eq:hsNonOscSolA}; this fixes either the scale factor exponent $b$ or
the nonminimal coupling parameter $\xi$. Now we proceed to study two of these
explicit examples.

\subsubsection{Explicit potential for fixed scale factor exponent}

The first explicit example is obtained by choosing the exponent as $b=-1/2$
and $\xi<1/4$, since \eqref{eq:hsNonOscSolA} becomes
\begin{equation}
\Psi(\tau)=\frac{\Psi_0}{\left[\left(\frac{\tau}{\tau_0}\right)^{\hat{\beta}}
+\left(\frac{\tau}{\tau_0}\right)^{-\hat{\beta}}
\right]^{\frac{1}{4\hat{\beta}^2}}},
\end{equation}
which is inverted as
\begin{equation}
\tau=\tau_0\left(\frac{1\pm\sqrt{1-4\left(\frac{\Psi}{\Psi_0}
\right)^{8\hat{\beta}^2}}}{2\left(\frac{\Psi}{\Psi_0}
\right)^{4\hat{\beta}^2}}\right)^{\frac{1}{\hat{\beta}}},
\end{equation}
where the dependence on the nonminimal coupling parameter is encoded in
\begin{equation}
\hat{\beta}=\sqrt{\frac{1-4\xi}{8\xi}}.
\end{equation}
Therefore, it is possible to obtain the two families of potentials
\begin{subequations}
\begin{align}
U_{\pm}(\Psi)={}&\frac{2^{\frac{1}{\hat{\beta}}}U_0}{16\hat{\beta}^2
(2\hat{\beta}^2+1)}\left(\frac{\Psi}{\Psi_0}\right)^{2(2\hat{\beta}+1)}
\frac{1}{\left[1\pm P(\Psi)\right]^{2+\frac{1}{\hat{\beta}}}}\nonumber\\
&\times\left[2(2\hat{\beta}+1)[3(2\hat{\beta}+1)\pm2(\hat{\beta}+1)P(\Psi)]
\rule{0pt}{6mm}\right.\nonumber\\
&\times
\left(\frac{\Psi}{\Psi_0}\right)^{8\hat{\beta}^2}
-8(2\hat{\beta}^2+1)\left(\frac{\Psi}{\Psi_0}\right)^{16\hat{\beta}^2}
\nonumber\\
&\left.{}-(2\hat{\beta}+1)(4\hat{\beta}+1)\left[1\pm P(\Psi)\right]
\rule{0pt}{6mm}\right],
\end{align}
with
\begin{equation}\label{eq:U02tau0Psi0}
P(\Psi)\equiv\sqrt{1-4\left(\frac{\Psi}{\Psi_0}\right)^{8\hat{\beta}^2}},\quad
U_{0}=\frac{\Psi_0\,^2}{16\tau_0a_0\,^2}.
\end{equation}
\end{subequations}
\begin{figure}
\includegraphics[width=\columnwidth]{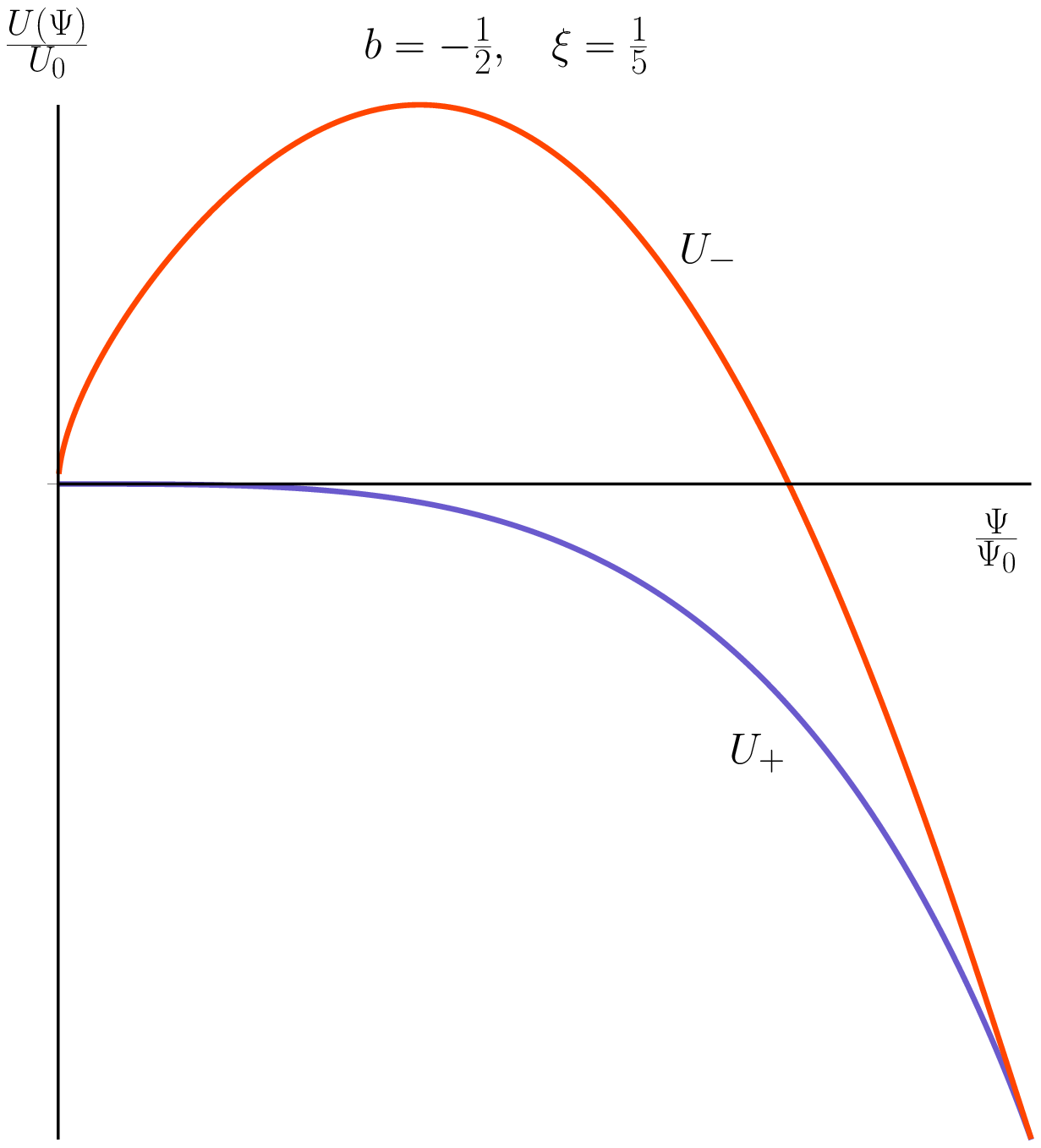}
\caption{The self-interaction potentials of the stealth overflying a power-law
universe with exponent $b=-1/2$ are plotted for $\xi=1/5$.}
\label{fig:PotPLhypES1}
\end{figure}%
As an illustration, the plot of the two branches $U_{+}(\Psi)$ and
$U_{-}(\Psi)$ with $\xi=1/5$ is presented in Fig.~\ref{fig:PotPLhypES1}.

\subsubsection{Explicit potential for fixed nonminimal coupling}

For the second explicit example, we fix the nonminimal coupling parameter in
terms of the exponent of the scale factor by
\begin{equation}\label{eq:tildxi}
\xi=\tilde{\xi}\equiv\frac{9b(b+1)}{4[13b(b+1)+1]},
\end{equation}
which cover the two general cases described at the beginning of this
subsection that give a positive discriminant \eqref{eq:discriminant}.
Concretely, in the first case the value $b=-1/2$ and the following intervals
are excluded
\begin{align}
-\frac12-\frac{3}{2\sqrt{13}}&<b<-\frac12-\frac{1}{\sqrt{6}}\nonumber,\\
-\frac12+\frac{1}{\sqrt{6}}&<b<-\frac12+\frac{3}{2\sqrt{13}}\nonumber,
\end{align}
since the defining condition $\tilde{\xi}<\xi_b$\ is only satisfied for the
rest of the interval $-1<b<0$. All the values of the second case are covered
by the parametrization \eqref{eq:tildxi}, because $\tilde{\xi}>\xi_b$ for
$b<-1$ and $b>0$.

After fixing the nonminimal coupling, expression \eqref{eq:hsNonOscSolA}
becomes
\begin{subequations}
\begin{align}
\Psi(\tau)&=\frac{\Psi_0}{\left[\left(\frac{\tau}{\tau_0}
\right)^{2\tilde{\beta}}
+\left(\frac{\tau}{\tau_0}\right)^{\tilde{\beta}}
\right]^{\frac{18\tilde{\beta}^2-1}{16\tilde{\beta}^2}}},
\label{eq:hsNonOscSolB}\\
\tilde{\beta}&=\frac{2b+1}{3}.\label{eq:betatilde}
\end{align}
\end{subequations}
For $b\neq-1/2$, this relation allows the inversion
\begin{equation}
\tau=\tau_0\left[\frac{-1+\sqrt{1+4\left(\frac{\Psi_0}{\Psi}
\right)^{\frac{16\tilde{\beta}^2}{18\tilde{\beta}^2-1}}}}{2}
\right]^{\frac{1}{\tilde{\beta}}}.
\end{equation}
Hence, the allowed potential is given by
\begin{subequations}
\begin{align}
U(\Psi)={}&\frac{2^{\frac{1}{\tilde{\beta}}+2}U_0(9\tilde{\beta}^2-1)^2}
{\tau_0\,^{6\beta}\tilde{\beta}^2(13\tilde{\beta}^2-1)}\left(\frac{\Psi}{\Psi_0}\right)^2
\left[Q(\Psi)-1\right]^{\frac{1}{\tilde{\beta}}-3}\nonumber\\
&\times\left\{[Q(\Psi)-1](13\tilde{\beta}^2-1)\left(\frac{\Psi}{\Psi_0}
\right)^{\frac{16\tilde{\beta}^2}{9\tilde{\beta}^2-1}}\right.\nonumber\\
&+2\left[2(13\tilde{\beta}^2+3\tilde{\beta}-1)Q(\Psi)-21\tilde{\beta}^2
-6\tilde{\beta}+3\right]\nonumber\\
&\times\left(\frac{\Psi}{\Psi_0}\right)^{\frac{8\tilde{\beta}^2}
{9\tilde{\beta}^2-1}}\left.-\frac{8(1+\tilde{\beta})
(3\tilde{\beta}^2+4\tilde{\beta}-1)}{3\tilde{\beta}+1}\right\}
\end{align}
where
\begin{equation}
Q(\Psi)\equiv\sqrt{1+4\left(\frac{\Psi}{\Psi_0}
\right)^{\frac{8\tilde{\beta}^2}{1-9\tilde{\beta}^2}}},
\end{equation}
\end{subequations}%
\begin{figure}
\includegraphics[width=\columnwidth]{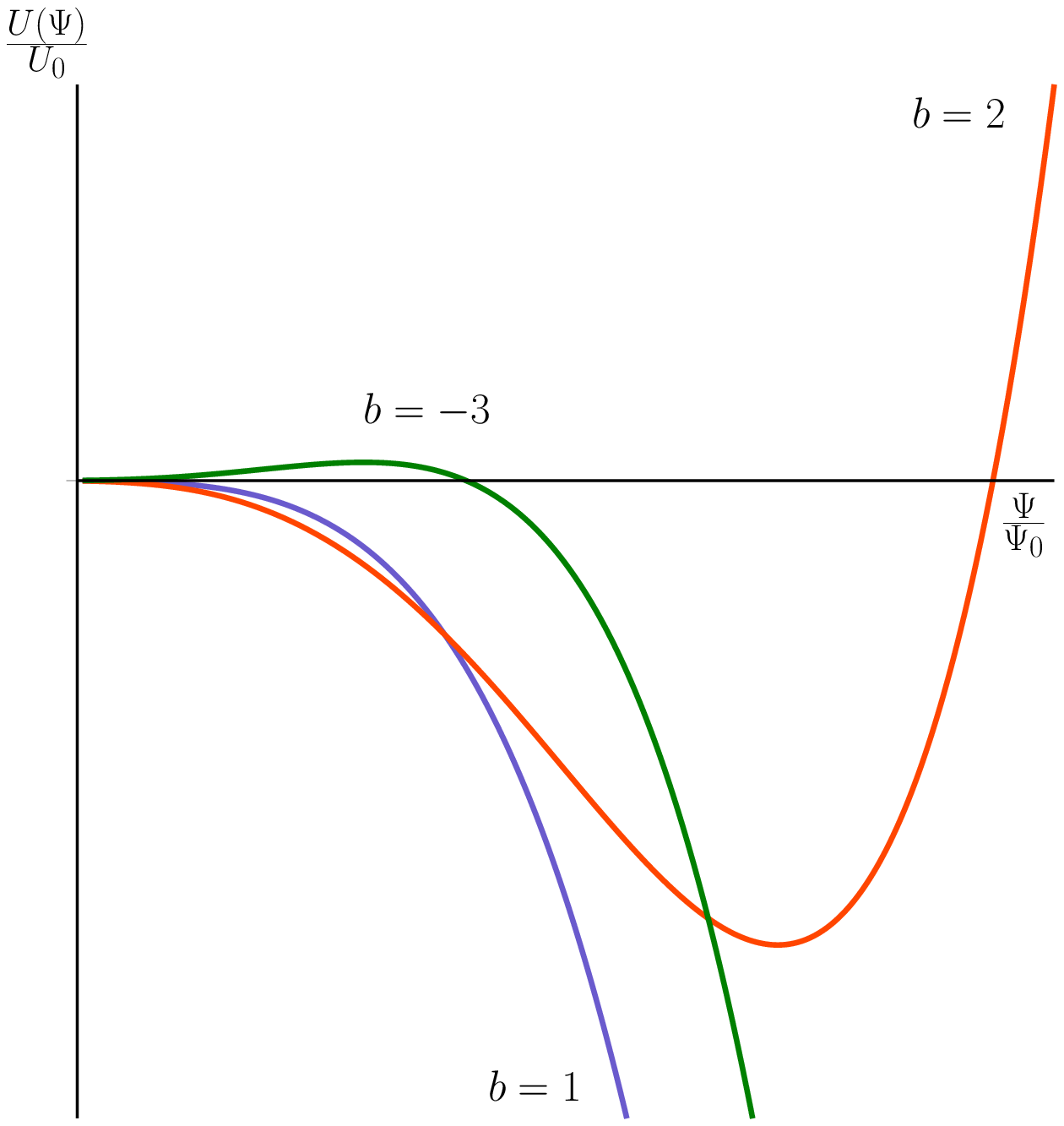}
\caption{The second example of an explicit self-interaction potential for the
stealth is plotted for radiation, $b=1$, matter, $b=2$, and inflation with
$b=-3$.}
\label{fig:PotPLhypES2RadMattInf}
\end{figure}%
and $U_0$ is still defined as in \eqref{eq:U02tau0Psi0}. In
Fig.~\ref{fig:PotPLhypES2RadMattInf} we plot the potentials for interesting
power laws corresponding to phases dominated by matter, radiation, and a case
of inflation.

\subsubsection{Early/late time approximation for arbitrary exponent
and coupling\label{subsubsec:e/l_time}}

For a generic phase characterized by a power-law exponent $b$ the stealth
potential cannot be explicitly obtained for an arbitrary value of the
nonminimal coupling parameter $\xi$. Nevertheless, we can approximate the
stealth solution for both early and late times in a given phase. The
resulting approximations turn out to be invertible giving the effective
potentials characterizing the self-interaction of the stealth in the
corresponding evolution regimes. Using the notations
$\Psi_-(\tau)=\Psi(\tau\ll\tau_0)$ and $\Psi_+(\tau)=\Psi(\tau\gg\tau_0)$
then both cases can be summarized within the same expression
\begin{equation}\label{eq:appxSol}
\Psi_{\pm}(\tau)=\Psi_0\left(\frac{\tau}{\tau_0}
\right)^{-\frac{\xi\left(2b+1\pm2\beta\right)}{1-4\xi}}.
\end{equation}
The explicit self-interaction potential for each approximation is
\begin{subequations}\label{eq:PotappxSol}
\begin{align}
U(\Psi_{\pm})=\lambda_{\pm}\left(\frac{\Psi_{\pm}}{\Psi_0}
\right)^{\frac{2[1-3\xi+(1-2\xi)b\pm2\xi\beta]}{\xi(2b+1\pm2\beta)}},
\end{align}
where the self-interaction coupling constants are given by
\begin{align}
\lambda_{\pm}={}&-\frac{16U_{0}\xi}{\tau_0\,^{2b+1}(4\xi-1)^2}
\left\{\pm2\xi[1+4(6\xi-1)b]\beta\right.\nonumber\\
&\left.{}+(6\xi-1)[2(8\xi-1)b+4\xi+1]b+\xi\right\}.
\end{align}
\end{subequations}%
\begin{figure}[t]
\includegraphics[width=\columnwidth]{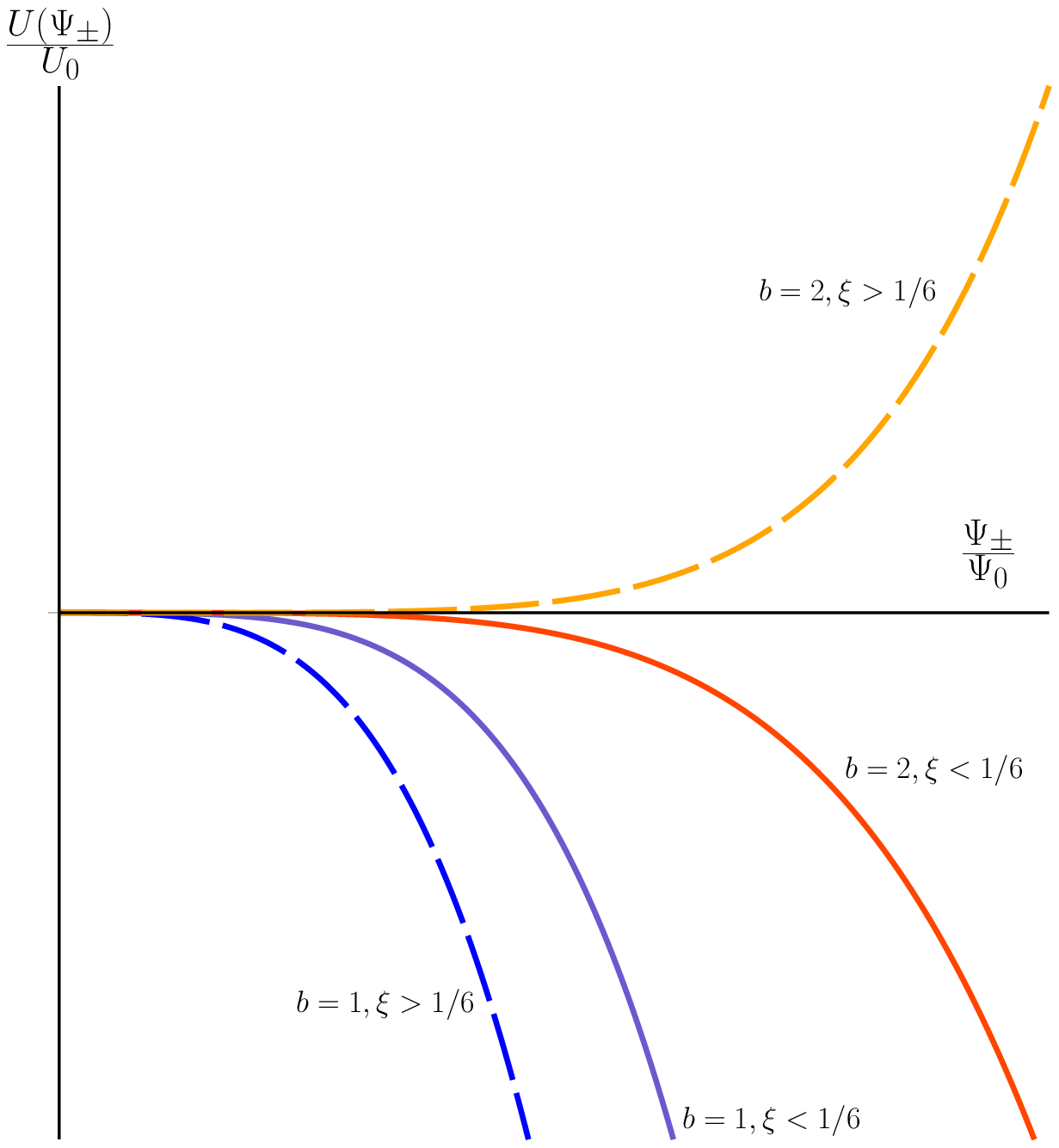}
\caption{Stealth self-interaction for different phases of the early universe
and values of the nonminimal coupling parameter $\xi$. The blue line,
$U(\Psi_+)$ corresponds to the late time radiation dominated epoch, $b=1$, and
the red line $U(\Psi_-)$ to the early matter dominated phase, $b=2$. The
graphs with $\xi<1/6$ are drawn with solid lines while for $\xi>1/6$ have long
dash lines.}
\label{fig:PotBBAppRadMatt}
\end{figure}%
We present in Fig.~\ref{fig:PotBBAppRadMatt} plots for two particular cases,
the late time approximation of the radiation dominated phase, $b=1$, and the
early time approximation of the matter dominated phase, $b=2$, for different
values of the nonminimal coupling parameter $\xi$. With these cases we end
the discussion of solutions with positive discriminant $\Delta$.

Summarizing the results in this subsection, we have found different explicit
solutions for the stealth problem when the discriminant
\eqref{eq:discriminant} is positive. We have obtained the self-interaction
potentials by fixing, on the one hand, the value of the scale factor exponent
and, on the other hand, the nonminimal coupling parameter in terms of the
former. We covered relevant phases of the Universe evolution, moreover, early
and late time regimes of these phases were also explicitly addressed. Several
behaviors of the potentials were found depending on the election of the
nonminimal coupling parameter and the scale factor exponent. Particularly, we
have found cases where the obtained potentials are unbounded from below,
however  this behavior can be mended as will be discussed in
Sec.~\ref{sec:Ueff}. In the following subsection we proceed to study
solutions with negative discriminant.

\subsection{Solutions with $\Delta<0$\label{subsubsec:xilxib}}

As it is shown in Fig.~\ref{fig:ExclusionZone}, the discriminant $\Delta$
given by \eqref{eq:discriminant} will be negative when $-1<b<0$ with
$\xi>\xi_b$, or in the case when $b>0$ or $b<-1$ with $\xi<\xi_b$. Therefore,
the power exponents of the solutions to the Euler equation
\eqref{eq:TEqBaroF} become imaginary, leading to trigonometric dependences.
Consequently, for $\Delta<0$ the stealth field can be expressed as
\begin{equation}	
\Psi(\tau)=\Psi_0\left[\left(\frac{\tau}{\tau_0}\right)^{b+1/2}\cos
\left(\sqrt{-\Delta}\ln\frac{\tau}{\tau_0}\right)
\right]^{-\frac{2\xi}{1-4\xi}}.
\end{equation}	
Once again, if $b=-1/2$ ($\xi>1/4$) it is possible to give an explicit
solution, since the relation between the stealth and the conformal time can
be inverted as
\begin{align}
\tau&=\tau_0\exp\left\{\frac{1}{\overline{\beta}}
\arccos\left[\left(\frac{\Psi}{\Psi_0}\right)^{4\overline{\beta}^2}
\right]\right\},\\
\overline{\beta}&=\sqrt{\frac{4\xi-1}{8\xi}}.
\end{align}
Therefore, the functional form of the potential is
\begin{figure}
\includegraphics[width=\columnwidth]{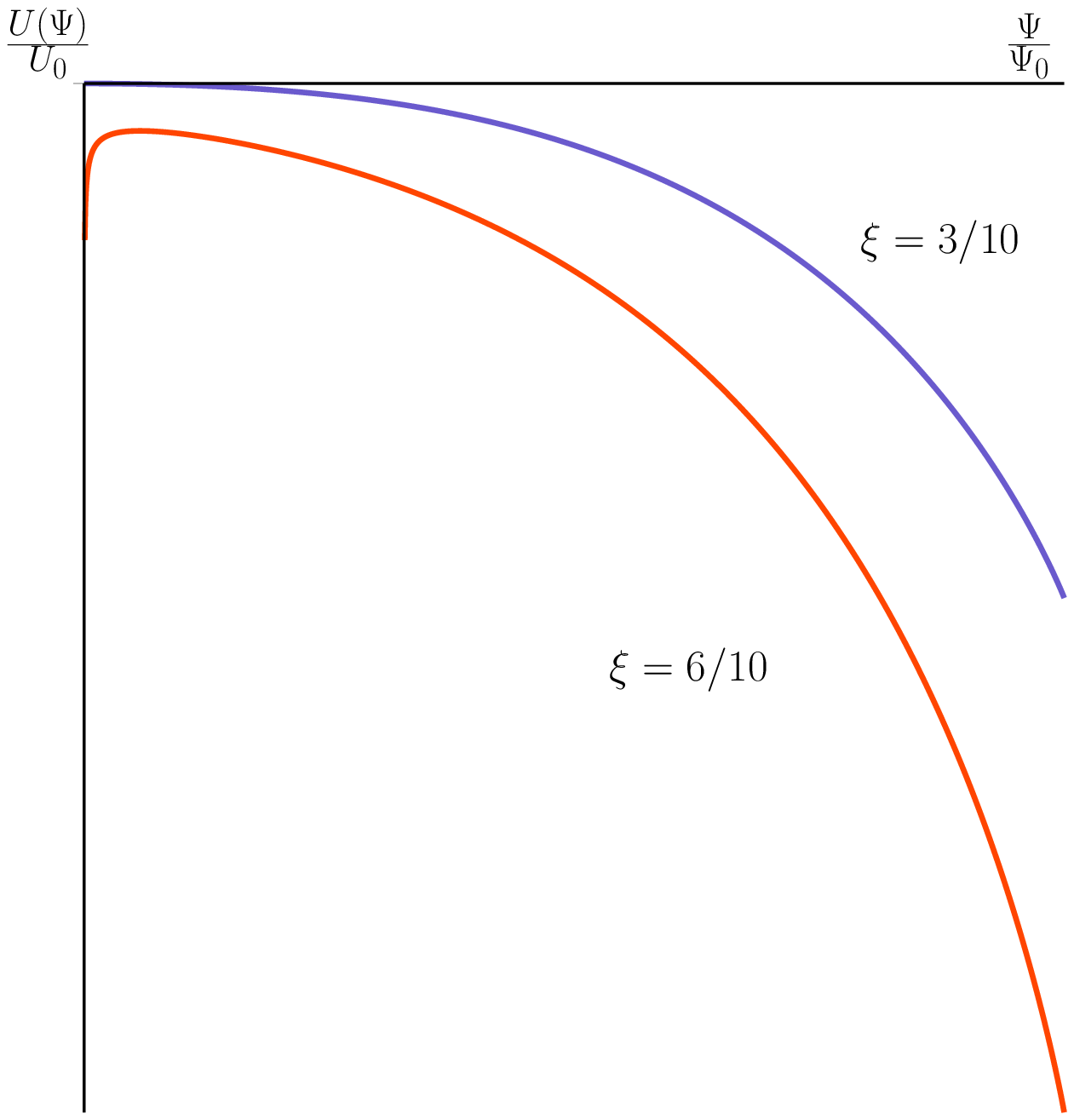}
\caption{Homogeneous stealth self-interaction of the oscillatory
solution with $b=-1/2$ for two values of $\xi>1/4$.}
\label{fig:PotPLoscES1}
\end{figure}
\begin{align}
U(\Psi)={}&4U_0\exp\left\{-\frac{1}{\overline{\beta}}
\arccos\left[\left(\frac{\Psi}{\Psi_0}\right)^{4\overline{\beta}^2}
\right]\right\}\left(\frac{\Psi}{\Psi_0}\right)^2\nonumber\\
&\times\left[\frac{1-2\overline{\beta}^2}{2\overline{\beta}^2}
\left(\frac{\Psi}{\Psi_0}\right)^{-8\overline{\beta}^2}
+\frac{1+2\overline{\beta}^2}{2\overline{\beta}^2}
\rule[2mm]{0pt}{8mm}\right.\nonumber\\
&+\left.\frac{3}{\overline{\beta}}
\sqrt{\left(\frac{\Psi}{\Psi_0}\right)^{-8\overline{\beta}^2}-1}\,
\rule[2mm]{0pt}{8mm}\right].
\end{align}
The plots of this potential are shown in Fig.~\ref{fig:PotPLoscES1}.

In this subsection we have studied the case were the discriminant
\eqref{eq:discriminant} is negative and, as in the previous subsection, an
explicit solution for the stealth and its self-interaction was found upon
fixing the scale factor exponent. It only remains to analyze the case with
vanishing discriminant in the next subsection.

\subsection{Solutions with $\Delta=0$}

We can appreciate from definition \eqref{eq:discriminant} that a vanishing
discriminant corresponds to the value of nonminimal coupling $\xi=\xi_b$,
implying that the Euler equation \eqref{eq:TEqBaroF} has a single root with
multiplicity two and the stealth can be found to be
\begin{equation}
\Psi(\tau)=\Psi_0\left[\left(\frac{\tau}{\tau_0}\right)^{b+\frac12}
\ln\frac{\tau}{\tau_0}\right]^\frac{-4b(b+1)}{(2b+1)^2}.
\end{equation}
Is not possible to obtain explicitly the potential in this case since this
relation is not invertible. Nevertheless, we can have an idea of the stealth
potential in presence of radiation or matter by plotting it parametrically
after fixing the corresponding exponent $b$. They exhibit a similar behavior
to the potential $U_{+}$ of Fig. \ref{fig:PotPLhypES1}.

Here, we exhaust the scanning of nonconformal stealths in universes with
power-law scale factor. Their necessarily homogeneous behavior is a
characteristic on all FRW universes, except for the de Sitter ones that we
study in the following section.

\section{de Sitter cosmologies with inhomogeneous stealths\label{sec:dS}}

In Sec.~\ref{sec:NCC} it was concluded that, except for the conformal
coupling $\xi=1/6$, the stealths of a generic universe must be necessarily
homogeneous, which are precisely the ones studied in the previous section.
The only exceptions allowing inhomogeneous stealths are the universes
satisfying the constraint \eqref{eq:dSConst}. As it is proved in
Appendix \ref{app:appdS} this constraint unambiguously defines the de Sitter
universes
\begin{equation}\label{eq:dSCosmologies}
ds^2=\frac{l^2k}{\sin^2\sqrt{k}\tau}\left(-d\tau^2+\frac{dr^2}{1-kr^2}
+r^2d\Omega^2\right).
\end{equation}
For them, even though the nonminimal coupling is not conformal, condition
\eqref{eq:T_te^te-T_t^t} lead us to the same time dependence obtained for the
stealth in the conformal case \cite{Ayon-Beato:2013bsa}, that is
\begin{equation}\label{eq:TdS}
T(\tau) =
\begin{cases}
\dfrac{A_0}{\sqrt{k}}\sin\sqrt{k}\tau
+B_{+}\cos\sqrt{k}\tau, & k\ne0,\\
-\frac12\alpha\tau^2+A_0\tau+\sigma_0, & k=0,
\rule{0pt}{6mm}
\end{cases}
\end{equation}
where $A_0$, $B_+$ and $\sigma_0$ are integration constants. This allows us
to write the full dependence of the auxiliary function characterizing the
stealth as
\begin{align}
\sigma(x^\mu)={}&\frac{A_0}{\sqrt{k}}\sin\sqrt{k}\tau
+\left(\frac{\sigma_0}2+\frac{\alpha}{k}\right)
\cos\sqrt{k}\tau\nonumber\\
&+\left(\frac{\sigma_0}2-\frac{\alpha}{k}\right)
\sqrt{1-k\,\vec{x}^2}+\vec{A}\cdot\vec{x}.\label{eq:sigmakpm1}
\end{align}
Here, the integration constants have been redefined as
$B_\pm=\sigma_0/2\pm\alpha/k$. Let us remind that the spatial integration
constants $A_1, A_2, A_3$ can be set to zero through quasitranslations as it
was done in Ref.~\cite{Ayon-Beato:2013bsa}. Finally, having the explicit
dependence of the stealth, we can find analytically the self-interaction
potential by solving constraint \eqref{eq:T00_const} which gives
\begin{subequations}
\begin{align}
U_{dS}(\Psi)={}&\frac{2\xi\Psi^2}{(1-4\xi)^2}
\Big[\xi\lambda_1(\sqrt{\kappa}\Psi)^\frac{1-4\xi}\xi \nonumber \\
&+24\lambda_2\xi (\xi-1/6) (\sqrt{\kappa}\Psi)^\frac{1-4\xi}{2\xi}
\nonumber \\
&-\frac{48}{l^2}(\xi-1/6)(\xi-3/16)\Big],
\label{eq:Pot_Inh}
\end{align}
where the coupling constants are defined by
\begin{align}
\lambda_1&=A_\mu A^\mu-2\alpha \sigma_0,\label{eq:lambda1}\\
\lambda_2&=A_0/l.\label{eq:lambda2}
\end{align}
\end{subequations}
This is exactly the self-interaction found in Ref.~\cite{AyonBeato:SAdS} for
the stealth on the dS background as is
reviewed in Appendix \ref{app:CompdS}.%
\begin{figure}
\includegraphics[width=\columnwidth]{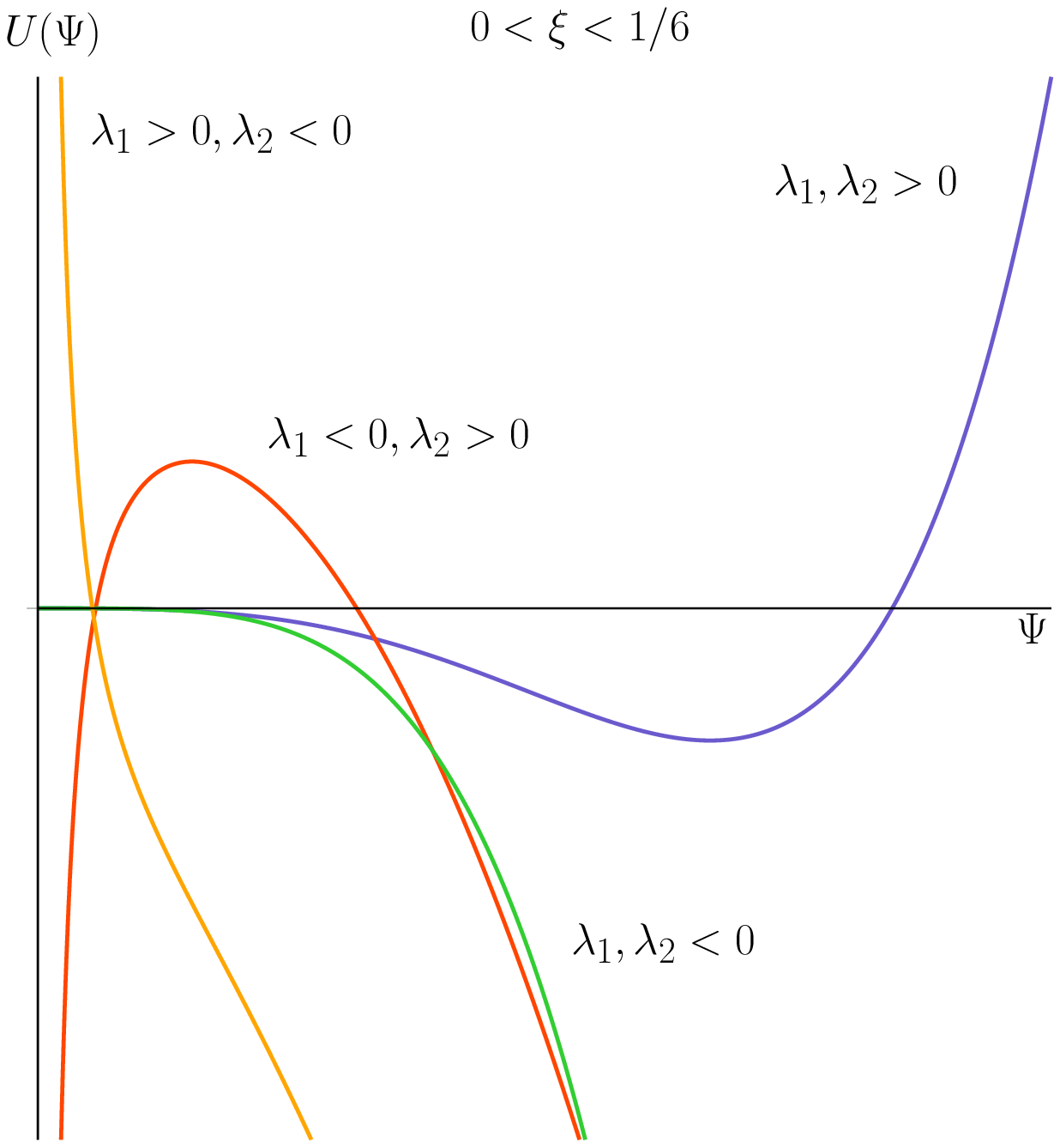}
\caption{Stealth self-interaction in de Sitter cosmologies for
different values of the coupling constants when $0<\xi<1/6$.}
\label{fig:dSxime16}
\end{figure}
\begin{figure}
\includegraphics[width=\columnwidth]{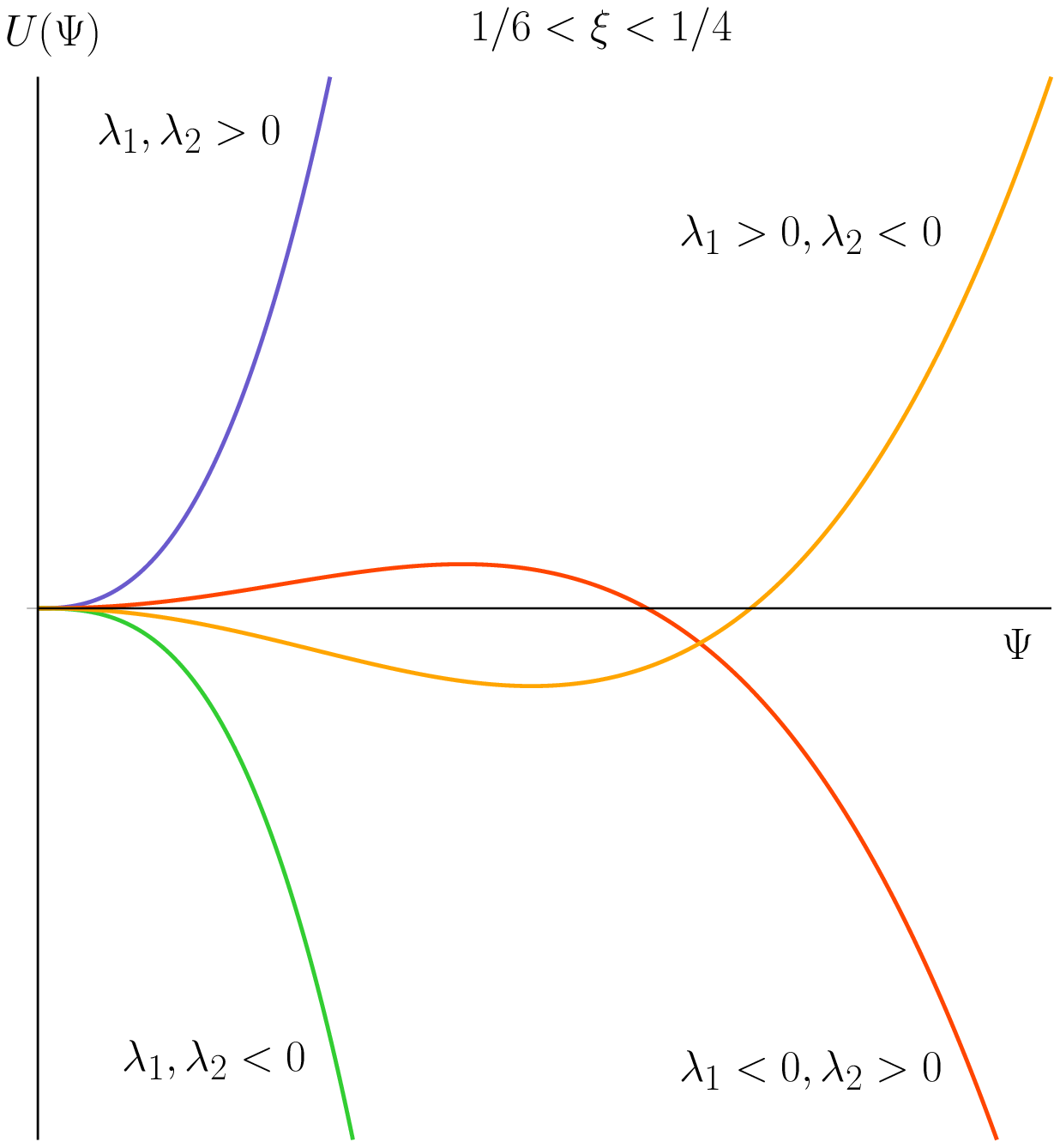}
\caption{Stealth self-interaction in de Sitter cosmologies for
different values of the coupling constants when $1/6<\xi<1/4$.}
\label{fig:dSxiMa16}
\end{figure}
\begin{figure}
\includegraphics[width=\columnwidth]{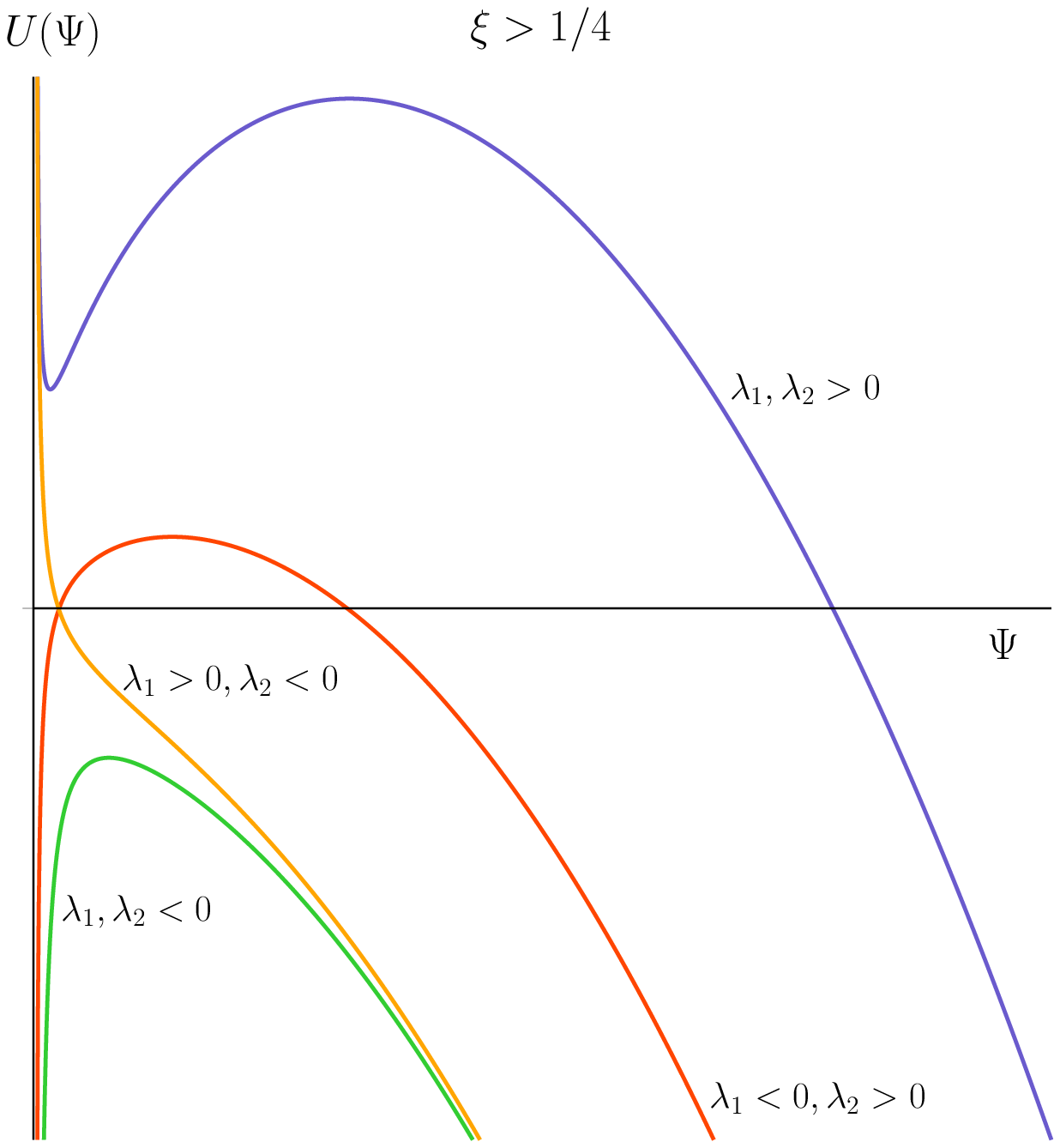}
\caption{Stealth self-interaction in de Sitter cosmologies for
different values of the coupling constants when $\xi>1/4$.}
\label{fig:dSCxiM14}
\end{figure}%
For $\xi = 1/6$ this potential becomes the conformal one already studied in
Ref.~\cite{Ayon-Beato:2013bsa} and exhibiting a single global minimum.
Whereas for $\xi\neq1/6$, the existence of their extrema depends on the
values of the coupling constants $\lambda_1$ and $\lambda_2$. For example, if
$\lambda_1>0$, $\lambda_2>0$ and $0<\xi<1/6$ the potential has a local
maximum at $\Psi=0$ and a global minimum (see Fig.~\ref{fig:dSxime16}).
Conversely, for $1/6<\xi<1/4$ there is only a global minimum at $\Psi=0$ for
the same range of coupling constants, as is shown in Fig.~\ref{fig:dSxiMa16}
along with other possible values for the couplings. Similarly, the cases with
$\xi>1/4$ are plotted in Fig.~\ref{fig:dSCxiM14} showing an unbounded from
below potential for all values of the couplings. However, these undesired
behaviors might be corrected when the nonminimal coupling contribution is
seen as part of the field self-interaction in its equation of motion. This
will be studied in detail in Sec.\ref{sec:Ueff}.

Up to now, all our analysis rest on the redefinition \eqref{eq:Psi2sigma}
which excludes the value $\xi=1/4$ of the nonminimal coupling. In the
following section we analyze how to handle this particular coupling.

\section{Nonminimal coupling $\xi=1/4$\label{sec:spcase14}}

Finally, in this section we address the atypical behavior of the stealth for
the coupling $\xi=1/4$, where its relation with the auxiliary function
\eqref{eq:Psi2sigma} is no longer valid. In this case the pertinent
redefinition is
\begin{equation}\label{eq:psi2asigma14}
\Psi=\frac{\exp{(a\sigma)}}{\sqrt{\kappa}},
\end{equation}
which allows the function $\sigma(x^\mu)$ to remain separable exactly the
same as it was before [see Eqs.~\eqref{eq:sigma_s} and \eqref{eq:RTP}]. This
way, the only dependence still to be determined is the temporal one, which is
rigged again by the combination \eqref{eq:T_te^te-T_t^t} which changes now to
\begin{align}
0&=-\frac{2a}{\Psi^2}\left(T_{~\theta}^{\mathrm{s}~\theta}
-T_{~\tau}^{\mathrm{s}~\tau}\right)\nonumber\\
&=
\begin{cases}
T''+kT
+\left(\mathcal{H}'-\mathcal{H}^2-k\right)\left(\sigma+\dfrac{1}{a}\right),
& k\ne0,\\ \\
T'' +\alpha
+\left(\mathcal{H}'-\mathcal{H}^2\right)\left(\sigma+\dfrac{1}{a}\right),
& k=0.
\end{cases}
\label{eq:T_te^te-T_t^t14}
\end{align}
As for any other coupling different from the conformal one, the solutions to
these equations are simplified for de Sitter cosmologies, giving rise to
inhomogeneous stealths to be studied in subsection \ref{sec:dS14}. For any other
cosmology the stealth has to be homogeneous, $\sigma(x^\mu)=T(\tau)$. In the
next subsection, motivated by the $\Lambda$CDM model, we will consider the
case of a power-law scale factor in a flat universe.

\subsection{Homogeneous stealth: power-law scale factor}

In a flat universe, $k=0$, with power-law scale factor $a(\tau)=a_0\tau^b$
the equations determining the temporal dependence of the stealth
\eqref{eq:T_te^te-T_t^t14} become again the Euler equation
\eqref{eq:TEqBaroF} for $\xi=1/4$ but with an additional power-law
inhomogeneity
\begin{equation}
\tau^2T''-b(b+1)T=\frac{b(b+1)}{a_0}\tau^{-b},
\end{equation}
whose solutions are now
\begin{equation}
T(\tau)=
\begin{cases}
A_0 \tau^{b+1}+\tau^{-b}\left(\sigma_0-\frac{b(b+1)}
{a_0(2b+1)}\ln\tau\right), & b\neq-\frac12,\\ \\
\sqrt{\tau}\left(A_0+\sigma_0\ln\tau-\frac{1}{8a_0}\ln^2\tau\right), &
b=-\frac12.
\end{cases}
\end{equation}

The corresponding stealth, which must be derived in these cases using the
redefinition \eqref{eq:psi2asigma14}, is not invertible for generic values of
the scale factor exponent. Nevertheless, again the exponent $b=-1/2$ is an
exception yielding
\begin{subequations}
\begin{equation}
\Psi(\tau)=\Psi_0\exp\left(-\frac18\ln^2\frac{\tau}{\tau_0}\right),
\end{equation}
where the integration constants are rewritten as
\begin{align}
\tau_0&=\exp(4a_0\sigma_0),\\
\Psi_0&=\frac{1}{\sqrt{\kappa}}\exp\left[a_0(A_0+2a_0\sigma_0^2)\right],
\end{align}
\end{subequations}
which allows to invert the conformal time as
\begin{equation}
\tau=\tau_0\exp\left(\pm2\sqrt{2\ln\frac{\Psi_0}{\Psi}}\right).
\end{equation}
Hence, the resulting self-interaction potential is
\begin{align}
U(\Psi)={}&-\frac{U_0\Psi^2}{\Psi_0^2}
\exp\left(-2\sqrt{2\ln\frac{\Psi_0}{\Psi}}\right)\nonumber\\
&\times\left(4\ln\frac{\Psi_0}{\Psi}+6\sqrt{2\ln\frac{\Psi_0}{\Psi}}+3\right),
\end{align}
where the constant $U_0$ is again given as in \eqref{eq:U02tau0Psi0}.
Potentials with similar qualitative behavior have been observed in previous
sections. For $b\neq-1/2$ the corresponding potentials can only be obtained
parametrically. In particular, this allows to verify that the plots for the
cases of matter and radiation are also similar to others presented in
previous figures.

\subsection{Inhomogeneous stealth: de Sitter cosmologies\label{sec:dS14}}

As it was emphasized after Eq.~\eqref{eq:T_te^te-T_t^t14}, inhomogeneous
stealths can only be present in de Sitter cosmologies. To obtain them for
$\xi=1/4$ we could proceed analogously as we did in Sec.~\ref{sec:dS} for the
other nonconformal couplings, i.e.\ to straightforwardly derive and solve the
equations for the temporal dependence of the stealth and to use this
knowledge to determine the potential. Nevertheless, for de Sitter universes
these equations are the same regardless of the value of $\xi$ and
consequently the auxiliary function $\sigma(x^\mu)$ remains unchanged. As it
was shown in Ref.~\cite{Ayon-Beato:2015qfa} the exponential dependence
present in redefinition \eqref{eq:psi2asigma14} for $\xi=1/4$ arises as a
nontrivial limit of the definition \eqref{eq:Psi2sigma} for the auxiliary
function corresponding to generic values of the coupling. We believe it is
more enlightening to exploit here this approach to find the $\xi=1/4$
potential as a nontrivial limit of \eqref{eq:Pot_Inh}.

\begin{figure}
\includegraphics[width=\columnwidth]{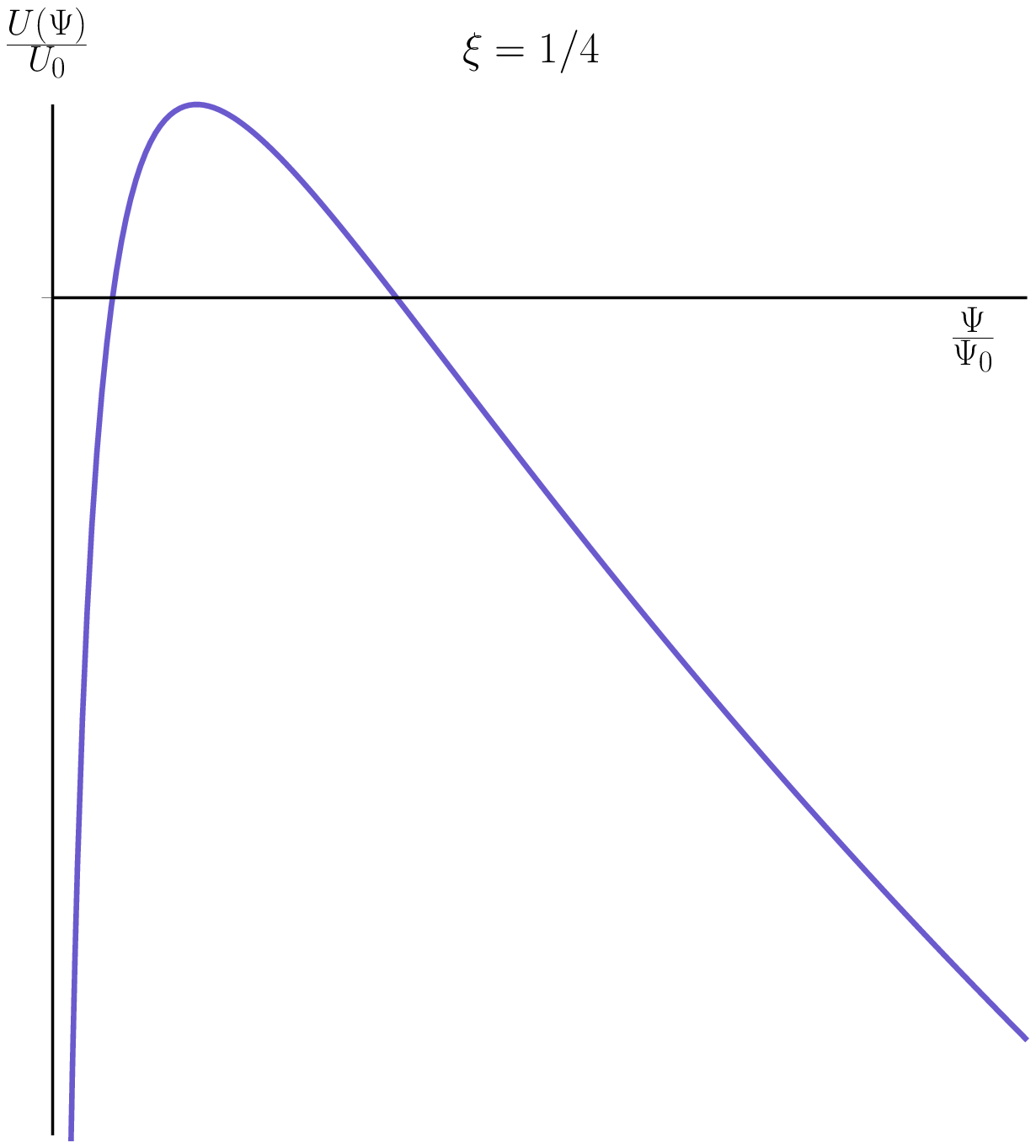}
\caption{Stealth self-interaction in de Sitter cosmologies when the nonminimal
coupling has the special value $\xi=1/4$.}
\label{fig:PotdSC14}
\end{figure}

The starting point of the procedure introduced in
Ref.~\cite{Ayon-Beato:2015qfa} is to consider the following limit
\begin{equation}\label{eq:limit1/4}
\lim_{\xi \to 1/4} \frac{2\xi(1-a(\tau)\sigma(x^\mu))}{1-4\xi}
\equiv a(\tau)\hat{\sigma}(x^\mu).
\end{equation}
If it is well behaved, modulo redefinitions of the integration constants, a
new auxiliary function could be defined according to the above right-hand
side. Then, starting from the expression for a generic coupling
\eqref{eq:Psi2sigma}, configuration \eqref{eq:psi2asigma14} can be reached as
the nonminimal coupling goes to the value $\xi=1/4$,  i.e.\
\begin{align}
\Psi&=\lim_{\xi \to 1/4}
\frac{1}{\sqrt{\kappa}}\left(a\sigma\right)^{-\frac{2\xi}{1-4\xi}}\nonumber\\
&=\lim_{\xi \to 1/4}
\frac{1}{\sqrt{\kappa}}\left(1-\frac{1-4\xi}{2\xi}a\hat{\sigma}
+\mathcal{O}\left((1-4\xi)^2\right)\right)^{-\frac{2\xi}{1-4\xi}}\nonumber\\
&=\frac{\exp(a\hat{\sigma})}{\sqrt\kappa}.
\end{align}
In the present case, the condition \eqref{eq:limit1/4} is achieved for the
stealth solution \eqref{eq:sigmakpm1} on de Sitter cosmologies
\eqref{eq:dSCosmologies} by redefining the integrations constants as
\begin{subequations}\label{eq:redctslimit14}
\begin{align}
A_0&=\frac1l-\frac{1-4\xi}{2\xi}\hat{A}_0,\\
\left(\sigma_0,\alpha,\vec{A}\right)&=-\frac{1-4\xi}{2\xi}\left(\hat{\sigma}_0,\hat{\alpha},\vec{\hat{A}}\right),
\end{align}
\end{subequations}
which allows to obtain exactly the same expression for the new auxiliary
function in terms of the new integration constants
\begin{align}\label{eq:sigmakpm11/4}
\hat{\sigma}(x^\mu)={}&\frac{\hat{A}_0}{\sqrt{k}}\sin\sqrt{k}\tau
+\left(\frac{\hat{\sigma}_0}2+\frac{\hat{\alpha}}{k}\right)
\cos\sqrt{k}\tau\nonumber\\
&+\left(\frac{\hat{\sigma}_0}2-\frac{\hat{\alpha}}{k}\right)
\sqrt{1-k\,\vec{x}^2}+\vec{\hat{A}}\cdot\vec{x}.
\end{align}
The next step is to obtain the $\xi=1/4$ potential as the limit resulting
from considering the redefinitions \eqref{eq:redctslimit14} on the starting
self-interaction \eqref{eq:Pot_Inh}, which after a careful rearrangement can
be rewritten as follows:
\begin{subequations}\label{eq:Pot14dS}
\begin{align}
U(\Psi)={}& \lim_{\xi \to 1/4}-\frac{\Psi^2}{2l^2}
\left[\left(1+\mathcal{O}(1-4\xi)\right)\rule{0pt}{7mm}\right.\nonumber\\
& \times\left(\frac{\Psi^{\frac{1-4\xi}{2\xi}}-1}{\frac{1-4\xi}{2\xi}}
-\hat{A_0}l+\frac{3}{2}+\mathcal{O}(1-4\xi)
	\right)^2\nonumber\\
&\left.\rule{0pt}{7mm} -\hat{\lambda}_1-\frac34+\mathcal{O}(1-4\xi)\right]
\nonumber\\
={}&-\frac{\Psi^2}{2l^2}\Bigg[\left(\ln\frac{\Psi}{\Psi_0}
+\frac32\right)^2-\hat{\lambda}_1-\frac34\Bigg],
\end{align}
where the new coupling constants are
\begin{align}
\hat{\lambda}_{1}&=l^2[\vec{\hat{A}}-2\hat{\alpha}\hat{\sigma}_0],\\
\Psi_0&=\frac{\exp(\hat{A}_0l)}{\sqrt{\kappa}}.
\end{align}
\end{subequations}
Expressions \eqref{eq:sigmakpm11/4} and \eqref{eq:Pot14dS} are just the exact
ones obtained by straightforwardly integrating the stealth constraints for
the nonminimal coupling $\xi=1/4$. These results are in agreement with those
previously found for (A)dS in Ref.~\cite{AyonBeato:SAdS}. The obtained
potential has extrema at
\begin{equation}
\Psi_{\pm}=\Psi_0\exp\left(-2\pm\sqrt{1+\hat{\lambda}_1}\right),
\end{equation}
which are local minima or maxima depending on the particular choices for
$\hat{\lambda}_1$. Different configurations are plotted in
Fig.~\ref{fig:PotdSC14}.

\section{Stealth effective potential \label{sec:Ueff}}
In previous sections we stated the problem of finding the stealth potential
in different cases. Particularly, we were able to determine the potential as
an explicit function of the field when the dependence of the stealth on the
conformal time was invertible. Nevertheless, it is important to notice that
to fully understand the stealth dynamics it must be considered that due to
the nonminimal coupling to gravity there is a term that can be considered as
part of an effective potential. Recall that the field equation of motion is
\begin{equation}\label{eq:ScFieEq}
\Box\Psi=U'(\Psi)+\xi R\Psi,
\end{equation}
which is obtained by taking the variation of action \eqref{eq:action+s} with
respect to the stealth. In a universe driven by a barotropic perfect fluid,
by taking the trace of the Einstein equations \eqref{eq:EEqs}, using the
conserved density \eqref{eq:dens_sf_rel} and the resulting power law
\eqref{eq:SFPL2} we obtain
\begin{equation}
R=\frac{2\kappa(b-1)}{b}\rho
=\frac{6b(b-1)}{a_b^2}\frac1{\tau^{2(b+1)}},
\end{equation}
i.~e.,~the Ricci scalar is proportional to the density of the fluid. As a
result, when the conformal time is invertible in terms of the stealth,
$\tau=\tau(\Psi)$, the scalar curvature can be expressed as a local function
of this scalar field, allowing to rewrite its equation of motion
\eqref{eq:ScFieEq} as follows
\begin{equation}\label{eq:ScFieEqEff}
\Box\Psi=U_\text{eff}'(\Psi).
\end{equation}
Hence, the effective potential is defined as
\begin{equation}\label{eq:U_eff}
U_\text{eff}(\Psi)\equiv U(\Psi)
+\frac{6{\xi}b(b-1)}{a_b^2}\int{\frac{\Psi}{\tau(\Psi)^{2(b+1)}} d\Psi}.
\end{equation}
Let us consider as nontrivial examples the particular cases of the early and
late time approximations studied in subsubsection \ref{subsubsec:e/l_time}. For
them, the inversion of their homogeneous stealths \eqref{eq:appxSol} in terms
of the conformal time gives power-law behaviors, which after being considered
in the integration of the nonminimal contribution of \eqref{eq:U_eff}, leads
to the same exponents appearing in their original potentials
\eqref{eq:PotappxSol}. Consequently the effective potentials are
\begin{subequations}\label{eq:PotappxSoleff}
\begin{equation}
U_\text{eff}(\Psi_{\pm})=\lambda_\pm^\text{eff}\left(\frac{\Psi_{\pm}}{\Psi_0}
\right)^{\frac{2[1-3\xi+(1-2\xi)b\pm2\xi\beta]}{\xi(2b+1\pm2\beta)}},
\end{equation}
where the correction is manifested via the related coupling constants
\begin{equation}
\lambda_{\pm}^{\text{eff}} \equiv \lambda_{\pm}
+\frac{48\xi^2(2b+1\mp2\beta)b(b-1)U_{0}}{[b+1-\xi(2b+3) \mp 2\xi\beta]\tau_{0}\,^{2b+1}}.
\end{equation}
\end{subequations}%
\begin{figure}
\includegraphics[width=\columnwidth]{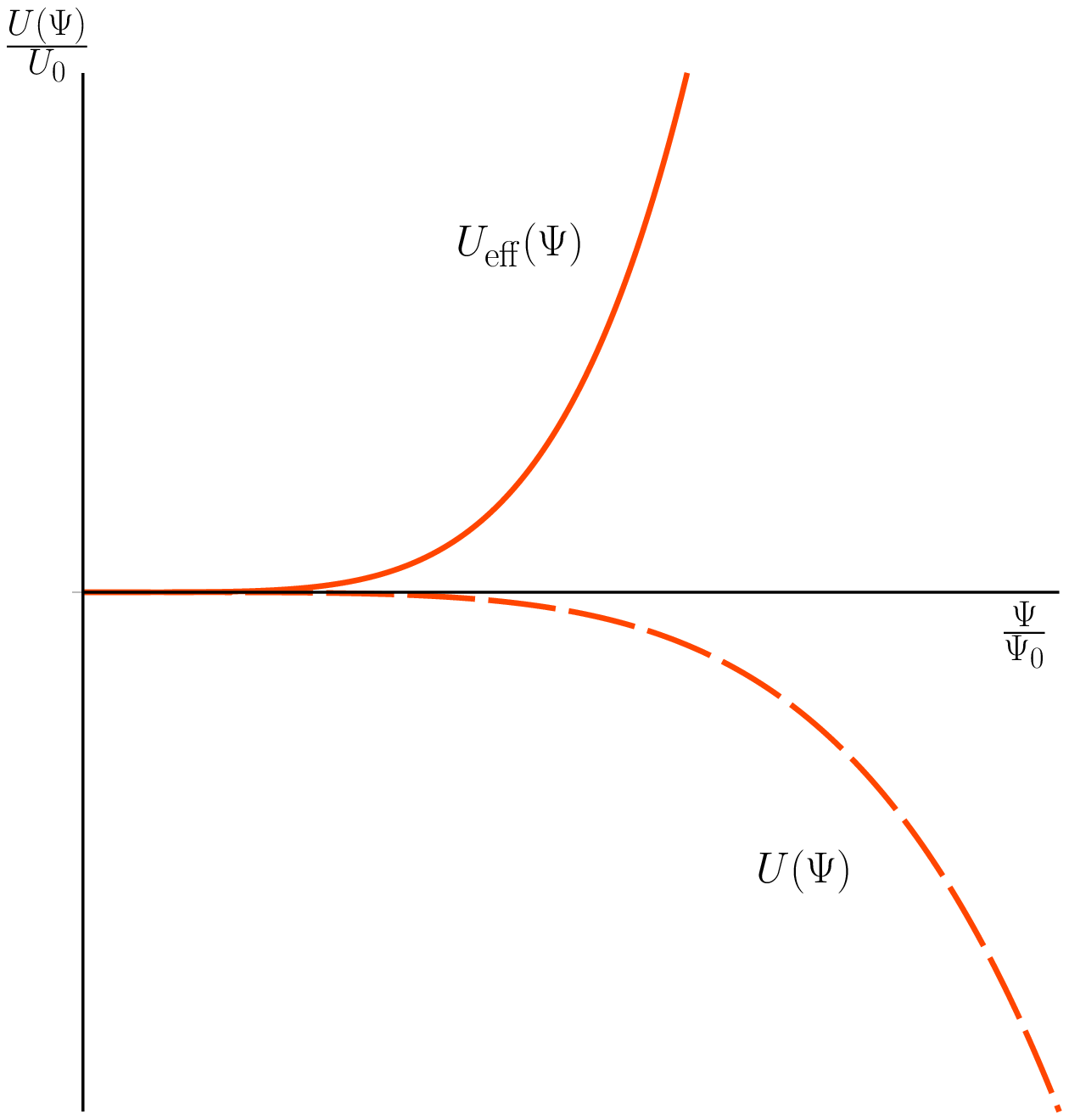}
\caption{The potential in presence of matter with $b=2$ is unbounded from
below when $\xi<1/6$, however considering the correction due to the nonminimal
coupling turns it well behaved.}
\label{fig:EffecPotMatt}
\end{figure}%
In some cases, like for instance stealths in presence of matter ($b=2$) with
a nonminimal coupling $\xi<1/6$, the self-interaction constants can change
its sign, drastically modifying the behavior of the potential as can be seen
in Fig.~\ref{fig:EffecPotMatt}.

\section{Conclusions and outlook}

The aim of the current work was to show that the $\Lambda$CDM model admits a
wider class of stealth fields than those conformally coupled to gravity
already studied in Ref.~\cite{Ayon-Beato:2013bsa}. The motivation behind this
plan is that most observables of this cosmological model are determined not
only by the time dependence of the scale factor but also by the evolution of
acoustic oscillations of density perturbations around the time when radiation
and matter densities were nearly equal. The introduction of the density
perturbations into the picture implies to break both the conformal flatness
of the FRW universe and the conformal symmetry of the stealth action
considered in \cite{Ayon-Beato:2013bsa}; the two indispensable ingredients
justifying the existence of the conformal stealths. Hence, the answer to the
question on the presence of stealths in the $\Lambda$CDM model necessarily
implies dealing with nonconformal couplings to gravity.

For more general nonminimal couplings we found in this research that
inhomogeneous stealths, as those enabled in the conformal case, are only
allowed in de Sitter backgrounds. Therefore, the $\Lambda$CDM stealth has to
be unavoidably homogeneous. We were able to find a version of such
nonconformal stealth by restricting the problem to a FRW universe filled with
radiation and matter. Unfortunately, the corresponding expression for this
stealth as function of the conformal time is given in terms of hypergeometric
functions, which makes it almost impossible to explicitly find the corresponding
self-interaction potential. The analogous task for the whole $\Lambda$CDM
model is even more challenging, but it could be approached numerically.
Moreover, an idea of how the potential for the $\Lambda$CDM stealth would
look can be drawn by solving the problem taking into account that in
each epoch of the cosmological evolution the corresponding scale factor has a
given power-law dependence on the conformal time. Here, we solved the problem
for homogeneous stealths during power-law inflation, radiation and matter
dominated eras, and during de Sitter expansion, which are the main epochs of
the $\Lambda$CDM model. We obtained a variety of functional forms for the
stealth potential, which may or may not be modified after considering the
contribution accounting for the nonminimal coupling to gravity in the stealth
equation of motion. For certain combinations of the nonminimal coupling $\xi$
and the scale-factor exponent $b$ this effective potential bounds the stealth
dynamics, in some cases allowing for the possibility of a stealth which
evolves towards a minimum in any of the main epochs of the $\Lambda$CDM
model.

There are two interesting and complementary ways of extending the results in
this paper to look for observational consequences of the existence of the
$\Lambda$CDM stealth. First of all, one can now consider the cases where
neither the stealth action is conformally invariant nor the spacetime is
conformally flat. This will allow us to establish whether a stealth field can
exist on inhomogenous or anisotropic universes like, for instance, several of
the Bianchi spacetimes. Complementarily, one can take the usual approach of
considering that the actual universe is well approximated by FRW plus small
perturbations. Using now the several examples of stealths found here and the
interesting fact that their fluctuations are not necessarily stealth
themselves, it would be possible to explore their consequences on the spectra
of the cosmic microwave background radiation and in the cosmological
large-scale structures statistics.

\begin{acknowledgments}
This research was supported by the Sistema Nacional de Investigadores
(Mexico). P.~I.~R.~-B was partially supported by "`Programa de Becas Mixtas"' from
CONACyT, by the “Plataforma de Movilidad Estudiantil Alianza del Pac\'ifico”
from AGCI and 15BEPD0003-II COMECyT grant. The work of C.~A.~T.~-E. was also
partially funded by FRABA-UCOL-14-2013 (Mexico). CAT-E also aknowledges the
warm hospitality at CINVESTAV-IPN during several visits when this research
was carried out.
\end{acknowledgments}

\appendix
	
\section{Scale factor for barotropic fluids in conformal time
\label{app:SF_BPF}}

As emphasized in the Introduction all cosmological phases can be modeled by
their dominant barotropic perfect fluid which imposes a power-law behavior
for the scale factor. Since this fact is intensively used the paper we
summarize here how it is described in terms of the conformal time. This allows
us to fix the notation for the remaining Appendixes and the whole paper.

A perfect fluid satisfies the continuity equation
\begin{equation}
\rho'+3	\mathcal{H}(\rho+P)=0.
\end{equation}
This fluid is denominated barotropic if additionally respects the equation of
state
\begin{equation}
P=w\rho,
\label{eq}
\end{equation}
which allows to integrate the continuity equation, yielding the following
relation between the scale factor and the density
\begin{subequations}\label{eq:dens_sf_rel}
\begin{align}
\rho&=\rho_0\left(\frac{a}{a_b}\right)^{-2(b+1)/b},
\end{align}
where
\begin{equation}
b=\frac{2}{3w+1},\qquad a_b=|b| \left(\frac{3}{\kappa\rho_0}\right)^{1/2},
	\label{eq:b2w}
\end{equation}
\end{subequations}
and the last is a length scale defined by the integration constant density
$\rho_0$. Using the above density \eqref{eq:dens_sf_rel} on the Friedmann
equation
\begin{equation}\label{eq:FriedmannEq}
3\mathcal{H}^2=\kappa\rho a^2-3k,
\end{equation}
for the flat case $k=0$ (which is the most interesting from the observational
point of view \cite{Aghanim:2015xee}) it becomes
\begin{equation}
\left(\frac{a'}{a_b}\right)^2=b^2\left(\frac{a}{a_b}\right)^{2(b-1)/b}.
\end{equation}
This equation is easily integrated yielding
\begin{equation}\label{eq:SFPL2}
a(\tau)=a_b \tau^b,
\end{equation}
modulo redefinitions of the conformal time making it positive in all cases.
This is the scale factor characterizing the conformal time evolution of a
barotropic perfect fluid and the starting point of the following two
Appendixes.

\section{Scale factor for a mixture of radiation and matter
\label{app:SF_mix_BPF}}

The first examples of stealths analyzed in Sec.~\ref{sec:LCDM} are those
present in the big bang era, which is characterized by the evolution of a
mixture of radiation ($w=1/3$) and pressureless matter ($w=0$). We explicitly
describe here the simultaneous evolution of these two kinds of barotropic
perfect fluids. Interestingly, in spite of the nonlinearity that Friedmann
equation \eqref{eq:FriedmannEq} inherits from Einstein equations, the
resulting scale factor reduces to a superposition of those each fluid obeys
separately. For a mixture of radiation, $b=1$, and matter, $b=2$,
\begin{equation}
\rho=\rho_\nu+\rho_{\mathrm{m}},
\end{equation}
since each component is individually conserved, their corresponding densities
\eqref{eq:dens_sf_rel} will be
\begin{align}
\rho_\nu=\rho_{\nu 0} \left(\frac{a}{a_1}\right)^{-4},\quad
\rho_{\mathrm{m}}=\rho_{\mathrm{m}0}\left(\frac{a}{a_2}\right)^{-3}.
\end{align}
Using these expressions, the flat Friedmann equation \eqref{eq:FriedmannEq}
reduces to
\begin{equation}
(a')^2=4a_2 a + (a_1)^2,
\end{equation}
which can be straightforwardly integrated as
\begin{equation}
a(\tau)=a_2(\tau-\tau_0)^2-\frac{(a_1)^2}{4a_2}.
\end{equation}
Surprisingly, after shifting the conformal time by
\begin{equation}
\tau\mapsto\tau-\tau_0-\frac{a_1}{2a_2},
\end{equation}
the scale factor for a mixture of radiation and matter can be written as a
superposition of the scale factors describing them individually
\begin{equation}
a(\tau)=a_1 \tau +a_2 \tau^2.
\end{equation}

\section{Power-law scale factors, cosmic vs conformal times
\label{app:SF_conf-cosm}}

Power-law behaviors on the scale factor are usually studied in terms of the
cosmic time. Since we use the conformal gauge in this paper, it is necessary
to reexamine the precise relation between both gauge choices. We address the
issue in this appendix. Recalling that the power-law scale factor in the
conformal time is given by \eqref{eq:SFPL2} then, from the relation between
both times
\begin{equation}\label{eq:tau2t}
dt=a d\tau,
\end{equation}
for $b\neq-1$ it is found that
\begin{equation}
t=\frac{a_b}{b+1}\tau^{b+1}.
\end{equation}
Therefore, the scale factor in the cosmic gauge is also given by a power law
%
\begin{equation}
a(t)=a_b\left(\frac{b+1}{a_b}t\right)^{\frac{b}{b+1}}\propto t^{p}, \qquad p\equiv \frac{b}{b+1}.
\end{equation}
The exception is the exponent $b=-1$ ($w=-1$), where the integration leads to
\begin{equation}
t=a_{-1}\ln \tau,
\label{eq:t2taum1}
\end{equation}
and consequently, the scale factor is described by the exponential behavior
\begin{equation}
a(t)=a_{-1}\exp\left(-\frac{t}{a_{-1}}\right).
\end{equation}
Notice from \eqref{eq:SFPL2} that $\tau<1$ in order to have a growing scale
factor, implying from \eqref{eq:t2taum1} that $t<0$. Hence, the usual
expression for de Sitter expansion in the cosmic gauge when $k=0$ (see
\eqref{eq:dSCcosmic} in the next Appendix) is obtained after a time
inversion.
	
\section{de Sitter cosmologies\label{app:appdS}}

Here, we briefly review vacuum Einstein equations with positive cosmological
constant for the FRW universes, which define de Sitter cosmologies, and how
they are equivalent to the constraint \eqref{eq:dSConst}. The cosmological
vacuum Einstein equations are diagonal as the metric
\begin{subequations}\label{eq:EE-dS-cosmo}
\begin{align}
G_\tau^{~\tau}+\Lambda&=
-\frac{3}{a^2}\left(\mathcal{H}^2+k\right)+\Lambda=0,
\label{eq:Ett}\\
G_{(i)}^{~~(i)}+\Lambda&=
-\frac{1}{a^2}\left(2\mathcal{H}'+\mathcal{H}^2+k\right)+\Lambda=0,
\label{eq:Eii}
\end{align}
\end{subequations}
where the index $i$ stands for the spatial components and the parenthesis
around it denotes the $i$-th diagonal mixed component and that no summation
occurs for this index. These equations are not independent, since the second
order differential equation resulting from spatial components \eqref{eq:Eii}
can be written in terms of the first order Friedmann one \eqref{eq:Ett} and
its derivative. Concretely, this is clear by taking their difference
\begin{align}
0=G_\tau^{~\tau}-G_{(i)}^{~~(i)}&=\frac{2}{a^2}\left(\mathcal{H}'
	-\mathcal{H}^2-k\right)\nonumber\\
&=\frac{1}{\mathcal{H}}\left(\frac{\mathcal{H}^2+k}{a^2}\right)'.
\label{eq:dS-cosmo}
\end{align}
As a result the concerned vacuum geometries necessarily obey the constraint
\eqref{eq:dSConst}. Conversely, if a given geometry fulfills constraint
\eqref{eq:dSConst} it also allows the existence of a first integral that can
be interpreted as the cosmological constant of a Friedmann-like equation. In
both situations it is enough to analyze the Friedmann equation
\eqref{eq:Ett}, whose solution is
\begin{equation}
a(\tau)=\frac{l\sqrt{k}}{\sin\sqrt{k}\tau},
\end{equation}
where $l=\sqrt{3/\Lambda}$ is the de Sitter radius associated to the involved
cosmological constant and the conformal time has been appropriately
reparametrized. This is the scale factor characterizing de Sitter
cosmologies in conformal time. The flat case is obtained as the limit
$k\rightarrow0$ of the above expression.

It is more common to express de Sitter cosmologies in terms of the cosmic
time. This is more easily obtained by straightforwardly solving the Friedmann
equation \eqref{eq:Ett} written in terms of this other gauge
\eqref{eq:tau2t}. The general result is
\begin{equation}
a(t)=l\left[\exp\left(\frac{t-t_0}{l}\right)+k\exp\left(-\frac{t-t_0}{l}\right)\right],
\end{equation}
which if splitted case by case gives, after appropriated reparametrizations,
the familiar expressions for de Sitter cosmologies
\begin{equation}\label{eq:dSCcosmic}
a(t)=
\begin{cases}
l\cosh t/l, & k=1,\\
l\exp t/l, & k=0,\rule{0pt}{5mm}\\
l\sinh t/l, & k=-1.\rule{0pt}{5mm}
\end{cases}
\end{equation}
	
\section{Stealth on de Sitter space from Riemann coordinates
\label{app:CompdS}}

The existence of stealths on de Sitter space was first studied in
Ref.~\cite{AyonBeato:SAdS} exploiting the fact that this space has constant
curvature, which implies it is also conformally flat. The last property is
manifest in the Riemann coordinates $x^\mu=(x^0,x^1,x^2,x^3)$ which allows us to
write de Sitter metric as
\begin{equation}\label{eq:dS_R}
ds^2=\Omega^2dx_\mu dx^\mu
\equiv\frac{dx_\mu dx^\mu}{\left(1+\frac{1}{4l^2}x_\mu x^{\mu}\right)^2}.
\end{equation}
In what follows the indices are lowered with the flat metric, e.g.\ $x_\mu
=\eta_{\mu\nu}x^\nu$. In Ref.~\cite{AyonBeato:SAdS} it is proved the stealth
is necessarily expressed by
\begin{subequations}\label{eq:S_R}
\begin{equation}
\Psi(x^\mu)= \frac{1}{\sqrt{\kappa}
\left(\Omega\sigma_\text{M}\right)^{\frac{2\xi}{1-4\xi}}},
\end{equation}
where $\Omega$ is the conformal factor of the metric \eqref{eq:dS_R} and the
auxiliary separable function corresponds to the one of Minkowski flat
spacetime \cite{AyonBeato:2005tu}
\begin{equation}\label{eq:s_R}
\sigma_\text{M}=
\frac{\tilde{\alpha}}2x_\mu x^\mu+\tilde{k}_\mu x^\mu
+\tilde{\sigma}_0,
\end{equation}
\end{subequations}
with $\tilde{\alpha}$, $\tilde{k}_\mu$ and $\tilde{\sigma}_0$ being
integration constants. The potential is exactly the same as
\eqref{eq:Pot_Inh} but the coupling constants are defined now in terms of the
present integration constants as
\begin{subequations}\label{eq:ccs_R}
\begin{align}
\lambda_1&=\tilde{k}_\mu \tilde{k}^\mu-2\tilde{\alpha}\tilde{\sigma}_0,\\
\lambda_2&=\tilde{\alpha}+\frac{\tilde{\sigma}_0}{2l^2}.
\end{align}
\end{subequations}

We shall show now that these results coincide with the ones we obtain in
Sec.~\ref{sec:dS} when de Sitter space is viewed as a FRW universe, i.e.\
when it is foliated by spatial hypersurfaces of constant curvature $k$. This
foliation can be accomplished by changing the Riemann coordinates to the
Cartesian FRW ones $(\tau,\vec{x})=(\tau,x,y,z)$, after substituting
\begin{subequations}\label{eq:Rm2FRW}
\begin{align}
x^0&=2l\frac{\left(\frac1k+\frac14\right)\cos\sqrt{k}\tau-\left(\frac1k-\frac14\right)\sqrt{1-k\vec{x}^2}}
{\frac{\sin\sqrt{k}\tau}{\sqrt{k}}+z},\\
x^1&=2l\frac{x}{\frac{\sin\sqrt{k}\tau}{\sqrt{k}}+z},\\
x^2&=2l\frac{y}{\frac{\sin\sqrt{k}\tau}{\sqrt{k}}+z},\\
x^3&=2l\frac{\left(\frac1k-\frac14\right)\cos\sqrt{k}\tau-\left(\frac1k+\frac14\right)\sqrt{1-k\vec{x}^2}}
{\frac{\sin\sqrt{k}\tau}{\sqrt{k}}+z}.
\end{align}
\end{subequations}
These transformations allow to express de Sitter metric \eqref{eq:dS_R} as
\begin{equation}
ds^2=\frac{kl^2}{\sin^2\sqrt{k}\tau}\left(-d\tau^2+d\vec{x}^2+\frac{k(\vec{x}
	\cdot d\vec{x})^2}{1-k\vec{x}^2}\right),
\end{equation}
which by passing to standard spherical coordinates looks in the form
\eqref{eq:dSCosmologies} used in Sec.~\ref{sec:dS}. Notice, the flat
foliation is obtained just by taking the limit $k\rightarrow0$ in
transformation \eqref{eq:Rm2FRW}. We start by writing the Riemann conformal
factor of \eqref{eq:dS_R} in the new coordinates
\begin{equation}
\Omega=\frac{\sin\sqrt{k}\tau+\sqrt{k}z}{2\sin\sqrt{k}\tau}.
\end{equation}
The next step is to change the Minkowski auxiliary function \eqref{eq:s_R} as
well
\begin{align}
\sigma_\text{M}={}&\frac{2\sqrt{k}l}{\sin\sqrt{k}\tau+\sqrt{k}z}\left.\rule{0pt}{6mm}\right\{\tilde{\alpha}l\left(\frac{\sin\sqrt{k}\tau}{\sqrt{k}}-z\right)\nonumber\\
&+\tilde{k}_0\left[\left(\frac{1}{k}+\frac14\right)\cos\sqrt{k}\tau-\left(\frac1k-\frac 14\right)\sqrt{1-kr^2}\right]\nonumber\\
&+\tilde{k}_1x+\tilde{k}_2y\nonumber\\
&+\tilde{k}_3\left[\left(\frac1k-\frac14\right)\cos\sqrt{k}\tau-\left(\frac1k+\frac14\right)\sqrt{1-kr^2}\right]\nonumber\\
&+\frac{\tilde{\sigma}_0}{2l}\left(\frac{\sin\sqrt{k}\tau}{\sqrt{k}}+z\right)\left.\rule{0pt}{6mm}\right\}\nonumber\\
={}&\frac{2\sqrt{k}l}{\sin\sqrt{k}\tau+\sqrt{k}z}\Bigg[
\frac{A_0}{\sqrt k}\sin\sqrt k\tau+\vec{A}\cdot\vec{x} \nonumber\\
&+\left(\frac{\sigma_{0}}{2}+\frac\alpha k\right)\cos\sqrt k\tau
\nonumber\\
&+\left(\frac{\sigma_0}{2}-\frac\alpha k\right)\sqrt{1-k\vec{x}^2}\Bigg],
\end{align}
where the integration constants are redefined by
\begin{subequations}\label{eq:ccR2ccFRW}
\begin{align}
A_0&=\frac{\tilde{\sigma}_0}{2l}+\tilde\alpha l, &
A_1&=\tilde{k}_1, &    \alpha&=\tilde{k}_0+\tilde{k}_3, \\
A_3&=\frac{\tilde{\sigma}_0}{2l}-\tilde\alpha l, &
A_2&=\tilde{k}_2, & 2\sigma_0&=\tilde{k}_0-\tilde{k}_3. 	
\end{align}
\end{subequations}
It is obvious now that the arguments of the powers in the redefinitions
\eqref{eq:S_R} and \eqref{eq:Psi2sigma} are exactly the same
\begin{equation}
\Omega\sigma_\text{M}=a\sigma_\text{FRW},
\end{equation}
and consequently, the expression for the stealth on de Sitter space written
in Riemann coordinates from Ref.~\cite{AyonBeato:SAdS} coincides with that
obtained in Sec.~\ref{sec:dS} when de Sitter is viewed as a cosmology.
Additionally, under relations \eqref{eq:ccR2ccFRW} the coupling constants of
the self-interaction potential \eqref{eq:ccs_R} are transformed to those
previously obtained in Eqs.~\eqref{eq:lambda1} and \eqref{eq:lambda2}.


\end{document}